\documentclass[%
 12pt,
 tightenlines,
 superscriptaddress,
 nofootinbib,
 notitlepage,
 bibnotes,
 amsmath,amssymb,
 aps,
 prd,
]{revtex4-1}
\usepackage{units}
\usepackage{graphicx}% Include figure files
\usepackage{dcolumn}% Align table columns on decimal point
\usepackage{bm}
\usepackage{leftidx}
\usepackage{bbold}
\usepackage[ruled,vlined]{algorithm2e}

\usepackage{amsfonts,amsmath,epsfig,bm,physics}
\usepackage[dvipsnames]{xcolor}
\usepackage{graphicx,hyperref}
\usepackage{booktabs}
\newcommand\numberthis{\addtocounter{equation}{1}\tag{\theequation}}
\usepackage{esvect}
\usepackage[caption=false]{subfig}
\hypersetup{
    colorlinks,
    citecolor=red,
    filecolor=red,
    linkcolor=Maroon,
    urlcolor=blue
}
\usepackage{mathtools}          %loads amsmath as well

\begin{document}

\preprint{APS/123-QED}

\title{Numerically modeling stochastic inflation in slow-roll and beyond}% Force line breaks with \\

\author{Aritra De}
\email{dexxx019@umn.edu}
\author{Rafid Mahbub}
\email{mahbu004@umn.edu}

\affiliation{%
 School of Physics \& Astronomy, University of Minnesota,\\
 Minneapolis, MN 55455, USA 
}%

\begin{abstract}
We present a complete numerical treatment of inflationary dynamics under the influence of stochastic corrections from sub-Hubble modes. We discuss how to exactly model the stochastic noise terms arising from the sub-Hubble quantum modes that give rise to the coarse-grained inflaton dynamics in the form of stochastic differential equations. The stochastic differential equations are solved event-by-event on a discrete time grid. We then compute the power spectrum of curvature perturbations that can be compared with the power spectrum computed in the traditional fashion using the Mukhanov-Sasaki equation by canonically quantizing the inflaton fluctuations. Our numerical procedure helps us to easily extend the formalism to ultra slow-roll inflation and study the possibility of primordial black hole formation.
\end{abstract}

\maketitle

\section{Introduction}

Cosmic inflation is one of the most robust theories in modern physics which is used to describe a phase of accelerated, quasi-de Sitter expansion in the very early history of the universe. Although it was first devised to explain the apparent shortcomings of the standard Big Bang picture \cite{Starobinsky:1980te,Guth:1980zm,Linde:1981mu}, inflation has turned out to be a theory with much more predictive power. We now know that quantum fluctuations of the inflaton get stretched to cosmological scales and give rise to metric and density perturbations and, hence, seed structure formation that are manifest in the CMB temperature fluctuations. \cite{Mukhanov:1981xt,1982PhLB..117..175S,Guth:1982ec,PhysRevD.28.679,1992PhR...215..203M}.\\
\indent One way to incorporate these quantum fluctuations is to split the inflaton into a background field, comprising of long wavelength modes, and small perturbative corrections which are comprised of short wavelength modes that initially start out inside the horizon and are later stretched out of causal contact due to the fact that, during inflation, the comoving Hubble radius $(aH)^{-1}$ decreases \cite{Starobinsky:1986fx}

\begin{equation}
\phi(t,\bm{x})=\bar{\phi}(t)+\hat{\phi}(t,\bm{x})
\end{equation}
where $\bar{\phi}$ refers to the classical, superhorizon inflaton field and $\hat{\phi}$ is the subhorizon, quantum part of the field that has not become classical. One can show that quantum-to-classical transitions can take place for cosmological perturbations and that the quantum operators can be replaced by stochastic noise terms that modify the inflaton evolution.\\
\indent The equation describing the quantum fluctuations is then recovered by demanding that the homogeneous component (background) vanishes due to the Klein-Gordon condition. In reality, the situation is not as straightforward and it can be shown that the quantum fluctuations backreact and modify the inflaton trajectory. This is precisely the physics that is captured by stochastic inflation which tracks the constant backreaction on the super-Hubble inflaton trajectory by the sub-Hubble quantum fluctuations as stochastic noise \cite{Starobinsky:1986fx,1994PhRvD..50.6357S,Tsamis:2005hd,Finelli:2008zg}.\\
\indent One of the main difficulties in the modeling of stochastic inflationary dynamics is that the numerical simulation of stochastic differential equations (SDEs) is both difficult and resource-intensive, even when SDEs are sourced by white noise processes. White noise is defined as a noise function with no correlation in space and time. Mathematically, it means that its two-point correlation is a Dirac $\delta$-function in space and time. This is the simplest form of noise implementation in stochastic inflationary dynamics. Even then, the exact modeling of such a noise term is computationally nontrivial and has not been adequately addressed in the existing literature, especially from the point of view of numerical modeling. An exact modeling of the noise (which is also relevant beyond slow-roll) involves tracking the evolution of the sub-Hubble quantum modes, which then influence the inflaton dynamics in a highly nontrivial manner. One of the primary objectives of this paper is to present the readers with a detailed description of how to solve these SDEs directly, which includes the modeling of noise terms without any slow-roll assumptions and simplifications. The benefit of this procedure is that it lets us calculate the noise amplitude in ultra slow-roll models (USR) as well. While most inflation models are slow-roll and a stochastic description can be obtained with the approximated noise, USR inflation has recently garnered some attention, especially in the domain of primordial black hole (PBH) formation \cite{Dimopoulos:2017ged,Dalianis:2018frf,Cicoli:2018asa,Garcia-Bellido:2017mdw}. The USR phase, which is basically a non-attractor phase, is usually characterized by a rapid deceleration of the inflaton field around an inflection point (more accurately, a \textit{near} inflection point\footnote{There have been debates regarding the conditions of an exact inflection point, i.e. $\partial_{\phi}V=\partial_{\phi\phi}V=0$ in the use of deriving model parameter sets. The choice of making $\partial_{\phi\phi}V\simeq 0$ has been discussed in \cite{Ezquiaga:2017fvi}.}) and deviations away from slow-roll give rise to non-negligible contributions from quantum fluctuations. In fact, an order of magnitude estimate reveals that the ratio of quantum fluctuations to classical drift goes as $\Delta\phi_{\text{qu}}/\Delta\phi_{\text{cl}}\sim\sqrt{\mathcal{P}_{\zeta}}$, where $\mathcal{P}_{\zeta}$ is the curvature power spectrum \cite{Pattison:2019hef}. It has been shown that the curvature power spectrum is amplified during the USR phase and, hence, quantum contributions dominate. One important consequence of increased quantum diffusion is that of an extra enhancement in the curvature power spectrum, which is typically not seen in classical calculations \cite{Ezquiaga:2018gbw}.\\
\indent Recently, the topic of solving SDEs using finite difference methods has been addressed in \cite{De:2020yyx} in the context of simulating stochastic hydrodynamic processes in heavy ion collisions. The authors have developed an approach to implementing random noise (as well as its derivatives and integrals) on a discretized spacetime grid. Such an implementation allows for the solutions of the SDEs to be computed on an  event-by-event basis. We will adopt a similar methodology when solving the SDEs in the current paper. Readers who are interested in the details of the finite difference schemes with noise terms and the appropriate interpretation of derivative and integral of noise terms on discrete lattices are referred to Ref. \cite{De:2020yyx}. \\
\indent In this paper, we first give a very brief introduction to inflationary dynamics in Sec. \ref{sec:inflationary_dynamics}. In Sec. \ref{sec:stochastic_inflation}, topics relevant to stochastic inflation are discussed which include exact expressions for noise terms and computation of the curvature power spectrum. In Sec. \ref{sec:numerical_solution_of_coarse_grained}, we dicuss the mathematics behind modeling the noise terms; a detailed exposition of the numerical techniques that have been developed to exactly model the stochastic noise and simulate the SDEs over a large number of realizations is presented in Sec. \ref{sec:numerical}. Finally, in Sec. \ref{sec:jackknife}, we address error estimates in our simulations using a technique called jackknife resampling on the power spectrum obtained from computing the SDEs over a large number of realizations. We work in units where $c=\hbar=1$ and the Planck mass $M_{\text{pl}}=(8\pi G)^{-1}$ is set to unity unless otherwise specified. The background cosmology is described by the Friedmann-Lema\^{i}tre-Robertson-Walker (FLRW) metric using the mostly positive metric convention: $ds^{2}=a(\eta)^{2}(-d\eta^{2}+\delta_{ij}dx^{i}dx^{j})$, where $\eta$ is the conformal time which is related to cosmic time $t$ through $dt=ad\eta$.

\section{Inflationary dynamics}\label{sec:inflationary_dynamics}
The simplest realization of the inflationary period is achieved by starting with the Einstein-Hilbert action and minimally coupling it to a scalar field $\phi$, called the inflaton.\footnote{Models with non-minimal coupling also exist but will not be discussed here. Interested readers are referred to \cite{Hertzberg:2010dc,Faraoni:1996rf}.} The action reads
\begin{equation}\label{eq:inflaton_action}
S=\int d^{4}x\sqrt{-g}\left[ \frac{1}{2}R-\frac{1}{2}g^{\mu\nu}\partial_{\mu}\phi\partial_{\nu}\phi -V(\phi) \right]
\end{equation}
where $g=\text{det}(g_{\mu\nu})$ is the determinant of the metric tensor and $R$ is the Ricci scalar $\phi$ is the inflaton field and $V(\phi)$ is the potential that drives inflation. One should note the relative signs on the kinetic and potential terms in the action. This is the sign convention in the Lagrangian density that arises due to the mostly positve metric convention. Using a mostly negative metric convention would produce a positive kinetic term. We consider the case where there is a single scalar field. Although we consider the case of a single scalar field, the physics can be generalized to accomodate the presence of multiple fields. The variation of the inflation action with respect to $\phi$ in Eq. \eqref{eq:inflaton_action} leads to the equation that describes the inflaton evolution
\begin{equation}
\frac{d^{2}\phi}{dN^{2}}+(3-\epsilon_{1})\frac{d\phi}{dN}+(3-\epsilon_{1})\frac{\partial_{\phi}V}{V}=0
\end{equation}
The inflaton evolution has been expressed in the $e$-fold time $N$ rather than cosmic time $t$ where the two are related by $dN=Hdt$. Here, $H$ is the Hubble parameter defined by

\begin{equation}
H^{2}=\frac{V(\phi)}{3-\frac{1}{2}\left( \frac{d\phi}{dN} \right)^{2}}
\end{equation}
The quantity $\epsilon_{1}$ is known as the first Hubble flow parameter
\begin{equation}
\epsilon_{1}=-\frac{1}{H}\frac{dH}{dN}
\end{equation}
The parameter $\epsilon_{1}$ is often used to track the progress of inflation since $\epsilon_{1}\ll 1$ throughout the duration of inflation, only reaching unity when inflation ends. Subsequent Hubble flow parameters are derived by the simple generalization
\begin{equation}
\epsilon_{n}=\frac{d\ln\epsilon_{n-1}}{dN}
\end{equation}
The practice that has been, more or less, standard is to split the inflaton into $\phi(N,\bm{x})=\bar{\phi}(N)+\delta\hat{\phi}(N,\bm{x})$ and, together with the Scalar-Vector-Tensor (SVT) decomposition of the FLRW metric \cite{Baumann:2009ds,Baumann:2018muz}, to introduce it to the Klein-Gordon equation describing inflaton. Once the equation of motion has been imposed on the homogeneous part of the field, one would end up with a description of the evolution of the fluctuating part, often called the Mukhanov-Sasaki equation. Details on this can be found in Appendix (\ref{sec:appendix_A}). In Fourier space, the Mukhanov-Sasaki equation is the following\footnote{The term $-2\epsilon_{1}(3-\epsilon_{1}+\epsilon_{2})$ can only be derived if the metric fluctuations are included. This is precisely what gives rise to the enhancement in $\mathcal{P}_{\zeta}$ in the USR period.}
\begin{equation}\label{eq:mukhanov_sasaki}
\frac{d^{2}\delta\phi_{k}}{dN^{2}}+(3-\epsilon_{1})\frac{d\delta\phi_{k}}{dN}+\left[ \left( \frac{k}{aH} \right)^{2} +(3-\epsilon_{1})\frac{\partial_{\phi\phi}V}{V}-2\epsilon_{1}(3-\epsilon_{1}+\epsilon_{2})\right]\delta\phi_{k}=0
\end{equation}
Equation \eqref{eq:mukhanov_sasaki} describes the evolution of the Fourier modes of the inflaton quantum fluctuations from an initially subhorizon regime ($k\gg aH$) to a superhorizon regime ($k\ll aH$). A common guideline for the demarcation of the subhorizon and superhorizon regimes is to consider $k=100a(N_{i})H(N_{i})$ and $k=0.01a(N_{f})H(N_{f})$, where $N_{i}$ and $N_{f}$ are the initial and final times of the evolution out of the total inflationary epoch which lasts $N$ $e$-folds. Inside the horizon the quantum modes do not feel the curvature of spacetime and the Bunch-Davies\cite{Bunch:1978yq} initial condition is imposed\footnote{This falls under the well known issue of finding a vacuum state for quantum field theories in curved spacetimes, which arises because the Hamiltonian is time-dependent. However, when quantum modes are inside the horizon, the spacetime appears Minkowski and a prescription for assigning a vacuum can be constructed. The Bunch-Davies vacuum is one such prescription which is typically defined as a zero particle state for geodesic observers. Readers interested in other kinds of vacua are directed to \cite{Mukhanov:2007zz} for details.} 
\begin{equation}
\delta\phi_{k}=\frac{1}{a\sqrt{2k}}\bigg\lvert_{N=N_{i}}\;\;\;\;\;\frac{d\delta\phi_{k}}{dN}=-\left( \frac{1}{a\sqrt{2k}}+i\frac{k}{aH}\frac{1}{a\sqrt{2k}} \right)\bigg\lvert_{N=N_{i}}
\end{equation}
The quantum fluctuations of the inflaton can then be used to define the gauge-invariant curvature perturbations\footnote{The notation $\zeta$ is a bit misleading since $\zeta$ is defined on uniform energy hypersurfaces. The definition in Eq. \eqref{eq:curvature_zeta} is really used to describe curvature perturbations on comoving hypersurfaces, denoted by $\mathcal{R}$. However, when $k\ll aH$, the two definitions coincide.}
\begin{equation}\label{eq:curvature_zeta}
\zeta_{k}=\Psi_{k} +\frac{\delta\phi_{k}}{d\bar{\phi} / dN}
\end{equation}
Here $\Psi_{k}$ is a metric scalar perturbation. It is described in detail in Appendix (\ref{sec:appendix_A}). A gauge-invariant quantity is one which does not change under a coordinate transfomation of the form $x^{\mu}\rightarrow x^{\mu}+\xi^{\mu}$. For the sake of convenience, one may choose a spatially flat gauge where $\Psi=0$, such that the curvature perturbations are given by $\zeta_{k}=\delta\phi_{k}/\sqrt{2\epsilon_{1}}$ and where $\epsilon_{1}=\frac{1}{2}(d\bar{\phi}/dN)^{2}$ follows from its usual definition. It can be shown that, on superhorizon scales, the curvature perturbations are conserved for adiabatic perturbations \cite{Lyth:2004gb,Riotto:2002yw} and the power spectrum of curvature perturbations are defined in this regime as follows
\begin{equation}
\mathcal{P}_{\zeta}(k)=\frac{k^3}{2\pi^2}|\zeta_{k}|^{2}_{k\ll aH}=\frac{k^3}{2\pi^2}\bigg\lvert \frac{\delta\phi_{k}}{\sqrt{2\epsilon_{1}}} \bigg\lvert^{2}_{k\ll aH}
\end{equation}
The background inflaton $\bar{\phi}$ tracks the progress of inflation while the quantum fluctuations $\delta\phi$ give rise to matter-energy fluctuations after reheating once they re-enter the horizon. Subsequently, we will see that an effective field theory description of the superhorizon inflaton field emerges with the influence of the subhorizon quantum modes appearing as stochastic noise. 

\section{Stochastic Inflation}\label{sec:stochastic_inflation}
In this section, we briefly discuss the stochastic inflation framework starting with the SDEs that govern the evolution of the coarse-grained inflaton field and the two-point correlation functions that characterize the noise. We also discuss methods to compute the power spectrum of curvature perturbations in this framework.
\subsection{Coarse-grained field evolution and noise correlation functions}
Stochastic inflation aims to provide an effective field theory framework that models the inflaton as a superhorizon, coarse-grained field that is being constantly modified by subhorizon quantum fluctuations. This effectively provides us with a classical, stochastic description of the inflaton evolution. To model such a system, the inflaton is split up as $\phi = \bar{\phi}+\delta\hat{\phi}$, where $\bar{\phi}$ is the superhorizon, coarse-grained field and $\delta\hat{\phi}$ describes the quantum fluctuations of the inflaton. The subhorizon fluctuations can be decomposed into a mode expansion as discussed earlier.
\begin{equation}
\delta\hat{\phi}(N,\bm{x})=\int_{k>0}\frac{d^{3}k}{(2\pi)^{3/2}}W\left( \frac{k}{\sigma aH} \right)e^{-i\bm{k}\cdot\bm{x}}\hat{a}_{\bm{k}}\delta\phi_{\bm{k}}(N)+\text{h.c.}
\end{equation}
where ``h.c." stands for the Hermitian conjugate of the mode expansion. The function $W(k/\sigma aH)$ is a suitably defined window function that picks out modes smaller than the horizon. As a result, the window function should be one such that $W\simeq 0$ when $k\ll \sigma aH$ and $W\simeq 1$ when $k\gg \sigma aH$. The coarse-graining scale is set by the parameter $\sigma\ll 1$. The value of $\sigma$ is chosen such that the coarse-graining scale is set well outside the Hubble horizon. In such a limit, the quantum nature of the inflaton fluctuations is lost due to decoherence and a quantum-to-classical transition justifies the use of the stochastic approach \cite{Lesgourgues:1996jc,Calzetta:1995ys,Mijic:1997mt,Kiefer:1998jk}. The nature of the stochastic process depends on the type of window function that has been used. The simplest and most commonly employed one is a sharp cut-off in momentum space
\begin{equation}\label{eq:sharp_window}
W\left( \frac{k}{\sigma aH} \right)=\Theta\left( \frac{k}{\sigma aH}-1 \right)
\end{equation}
This type of window function produces noise that is uncorrelated in time (white noise). To understand why this happens, consider the fact that in the definition of the noise terms the window function appears as a time derivative \cite{Vennin:2014lfa}
\begin{align*}
\frac{\partial}{\partial N}W\left( \frac{k}{\sigma aH} \right)&=\frac{k}{\sigma aH}(\epsilon_{1}-1)W'\left( \frac{k}{\sigma aH} \right)\\
&=k(\epsilon_{1}-1)\delta(k-\sigma aH)\numberthis 
\end{align*}
A more physically motivated choice for a window function may be a Gaussian one. However, more complicated window functions tend to produce colored noise where the noise terms are correlated in time \cite{Winitzki:1999ve,Liguori:2004fa}. A numerical implementation of SDEs with colored noise is deferred to a future work.\\
\indent The stochastic evolution of the coarse-grained inflaton field is usually studied by introducing a canonical momentum field $\bar{\pi}_{\phi}=d\bar{\phi}/dN$ and splitting the inflaton evolution into two coupled first-order stochastic differential equations of the form
\begin{align*}\label{eq:coarse_grain_evol}
\frac{d\bar{\phi}}{dN}&=\bar{\pi}_{\phi}+\xi_{\phi}\\
\frac{d\bar{\pi}_{\phi}}{dN}&=-(3-\epsilon_{1})\left( \bar{\pi}_{\phi}+\frac{\partial_{\phi}V}{V} \right)+\xi_{\pi}\numberthis
\end{align*}
where $\xi_{\phi}$ and $\xi_{\phi}$ are the noise terms associated with the coarse-grained inflaton field and its conjugate momentum. Details of this derivation can be found in Appendix (\ref{sec:appendix_B}). The correlation functions of the noise terms can be found in \cite{Vennin:2014lfa,Grain:2017dqa}. The statistical properties of $\xi_{\phi}$ and $\xi_{\pi}$ are encoded in a correlation function matrix of the following form
\begin{equation}
\bm{\Xi}(\bm{x}_{1},N_{1};\bm{x}_{2},N_{2})=\mqty(\langle \xi_{\phi}(\bm{x}_{1},N_{1})\xi_{\phi}(\bm{x}_{2},N_{2}) \rangle&\langle \xi_{\phi}(\bm{x}_{1},N_{1})\xi_{\pi}(\bm{x}_{2},N_{2}) \rangle\\\langle \xi_{\pi}(\bm{x}_{1},N_{1})\xi_{\phi}(\bm{x}_{2},N_{2}) \rangle&\langle \xi_{\pi}(\bm{x}_{1},N_{1})\xi_{\pi}(\bm{x}_{2},N_{2}) \rangle)
\end{equation}
The matrix elements of $\bm{\Xi}$ may be labelled as $\Xi_{f,g}$ where $f$ and $g$ are either $\phi$ and (or) $\pi_{\phi}$. For a general window function, the correlation function matrix reads
\begin{multline}\label{eq:correlation_matrix}
\bm{\Xi}(\bm{x}_{1},N_{1};\bm{x}_{2},N_{2})=\int\frac{d^{3}k}{(2\pi)^{3}}\frac{\partial}{\partial N}W\left( \frac{k}{k_{\sigma}(N_{1})} \right)\frac{\partial}{\partial N}W\left( \frac{k}{k_{\sigma}(N_{2})} \right) \\
\times e^{i\bm{k}\cdot(\bm{x}_{2}-\bm{x}_{1})}\mqty(\delta\phi_{k}(N_{1})\delta\phi_{k}^{*}(N_{2})&\delta\phi_{k}(N_{1})\delta\pi_{k}^{*}(N_{2})\\\delta\pi_{k}(N_{1})\delta\phi_{k}^{*}(N_{2})&\delta\pi_{k}(N_{1})\delta\pi_{k}^{*}(N_{2}))
\end{multline}
where $k_{\sigma}(N)=\sigma a(N)H(N)$. This is the most general expression for the correlation function matrix and can be used to describe any suitable window function. In the case of the window function defined in Eq. \eqref{eq:sharp_window}, the correlation functions simplify to terms proportional to Dirac $\delta$-functions. In the end, it can be shown that the correlation functions reduce to
\begin{align*}\label{eq:noise_correlation}
\Xi_{fg}(\bm{x}_{1}-\bm{x}_{2};N_{1}-N_{2})&=\frac{k_{\sigma}^{3}(N_{1})}{2\pi^2}(1-\epsilon_{1}(N))f_{k=k_{\sigma}(N_{1})}g_{k=k_{\sigma}(N_{1})}^{*}\frac{\sin\left[ k_{\sigma}(N_{1})|\bm{x}_{2}-\bm{x}_{1}| \right]}{k_{\sigma}(N_{1})|\bm{x}_{2}-\bm{x}_{1}|}\delta(N_{1}-N_{2})\\
&=(1-\epsilon_{1}(N))\mathcal{P}_{fg}(k_{\sigma})\frac{\sin\left[ k_{\sigma}(N_{1})|\bm{x}_{2}-\bm{x}_{1}| \right]}{k_{\sigma}(N_{1})|\bm{x}_{2}-\bm{x}_{1}|}\delta(N_{1}-N_{2})\numberthis
\end{align*}
where $\mathcal{P}_{fg}$ is the dimensionless power spectrum of the form $fg^{*}$ evaluated at $k_{\sigma}$. For example, the term $\Xi_{\phi\phi}$ will be governed by the power spectrum of inflaton fluctuations $\mathcal{P}_{\delta\phi\delta\phi}$. Equation \eqref{eq:noise_correlation} is the exact expression for the noise correlation functions of $\delta\phi$ and $\delta\pi_{\phi}$ and their cross correlations. A complete numerical treatment involves evaluating the power spectrum at each time step for the corresponding Fourier mode $k_{\sigma}$. In slow-roll approximation, $\epsilon_{1}\simeq 0$ and the mode functions (in the superhorizon regime) are given by \cite{Riotto:2002yw,Ezquiaga:2018gbw}
\begin{equation}\label{eq:modes_SR}
\delta\phi_{k}=\frac{H}{\sqrt{2k^{3}}}\left(\frac{k}{aH} \right)^{\frac{3}{2}-\nu}\;\;\;\;\;\delta\pi_{k}=\frac{H}{\sqrt{2k^{3}}}\left( \nu-\frac{3}{2} \right)\left( \frac{k}{aH} \right)^{\frac{3}{2}-\nu}
\end{equation}
where $\nu^{2}=\frac{9}{4}-\frac{m^{2}}{H^{2}}$. A standard approach in the slow-roll calculation is one where a massless scalar field in de Sitter space is considered for which $\nu\sim3/2$. Using the mode expansions, we can compute the $\phi$ and $\pi$ correlation functions in a straightforward manner. With the definition of dimensionless power spectrum and considering correlations at equal spatial points,
\begin{align*}\label{eq:correlations_SR}
\Xi_{\phi\phi}&=\frac{k_{\sigma}^{3}}{2\pi^2}\frac{H^{2}}{2k_{\sigma}^{3}}\left( \frac{k_{\sigma}}{aH} \right)^{\frac{3}{2}-\nu}=\frac{H^{2}}{4\pi^{2}}\sigma^{3-2\nu}\\
\Xi_{\pi\pi}&=\frac{k_{\sigma}^{3}}{2\pi^2}\frac{H^{2}}{2k_{\sigma}^{3}}\left( \nu-\frac{3}{2} \right)^{2}\left( \frac{k_{\sigma}}{aH} \right)^{\frac{3}{2}-\nu}=\frac{H^{2}}{4\pi^{2}}\left( \nu-\frac{3}{2} \right)^{2}\sigma^{3-2\nu}\numberthis
\end{align*}
From here on the notation we shall adopt is one where the correlation functions are given as $\langle \xi_{f}(\bm{x}_{1},N_{1})\xi_{g}(\bm{x}_{2},N_{2}) \rangle\equiv\Xi_{fg}(\bm{x}_{1}-\bm{x}_{2};N_{1})\delta(N_{1}-N_{2})$ where $\Xi_{fg}$ simply encodes the amplitude of the correlation functions.

In the massless de Sitter limit, $\Xi_{\phi\phi}\simeq H^{2}/4\pi^{2}$ and $\Xi_{\pi\phi}\simeq 0$. This is a usable approximation, at least in the slow-roll case, since $m^{2}/H^{2}=\partial_{\phi\phi}V/V\equiv\eta$, where $\eta$\footnote{Here $\eta$ is defined as a potential slow-roll parameter because the definition relies on the inflaton potential and its derivatives. More general expressions for these come in the form of the Hubble flow parameters discussed in the previous section.} is a slow-roll parameter which remains small throughout inflation. In more general settings, Eq. \eqref{eq:modes_SR} and subsequently \eqref{eq:correlations_SR} cannot be used for accurate calculations primarily because the mode expansions work only in the slow-roll limit where $H$ is assumed to be constant.\footnote{This is technically not true even in slow-roll since the inflationary phase is really a period of quasi de Sitter expasion. A perfect de Sitter expansion would have $H=\text{constant}$.}  However, for more non-trivial inflation models, especially ones possessing a USR period, the $\Xi_{\pi\pi}$ term becomes of the order of $\Xi_{\phi\phi}$ around the plateau region and cannot be ignored.

\subsection{Power spectrum of curvature perturbations}\label{sec:power}
It should be emphasized that the usual prescription for calculating the curvature perturbations, by initiating the inflaton fluctuations deep inside the horizon and evolving them, does not apply in the stochastic inflation framework. It is due to the fact that quantum fluctuations have been replaced by classical noise and it makes sense to talk about the inflaton only at the coarse-graining scale \cite{Levasseur:2014ska}. Nevertheless, the curvature perturbations, and the power spectrum thereof, can be computed with a different line of interpretation. We recall that, in the spatially flat gauge, the curvature perturbations are defined as 
\begin{equation}
\zeta_{k}=\frac{\delta\phi_{k}}{\sqrt{2\epsilon_{1}}}
\end{equation} 
where $\delta\phi_{k}$ are the Fourier modes of the inflaton fluctuations. We also know that an alternative definition of $\mathcal{P}_{\zeta}$ comes from the two-point function of $\zeta(\bm{x})$
\begin{equation}
\langle \zeta(\bm{x})^{2} \rangle=\int\frac{dk}{k}\mathcal{P}_{\zeta}(k)
\end{equation}
such that 
\begin{equation}\label{eq:P_zeta_def}
\mathcal{P}_{\zeta}(k)=\frac{d\langle \zeta^{2} \rangle}{d\ln k}=\frac{d}{d\ln k}\left( \frac{\langle \delta\phi_{k}^{2} \rangle}{2\epsilon_{1}} \right)
\end{equation}

\begin{figure}
\centering
\includegraphics[scale=0.8]{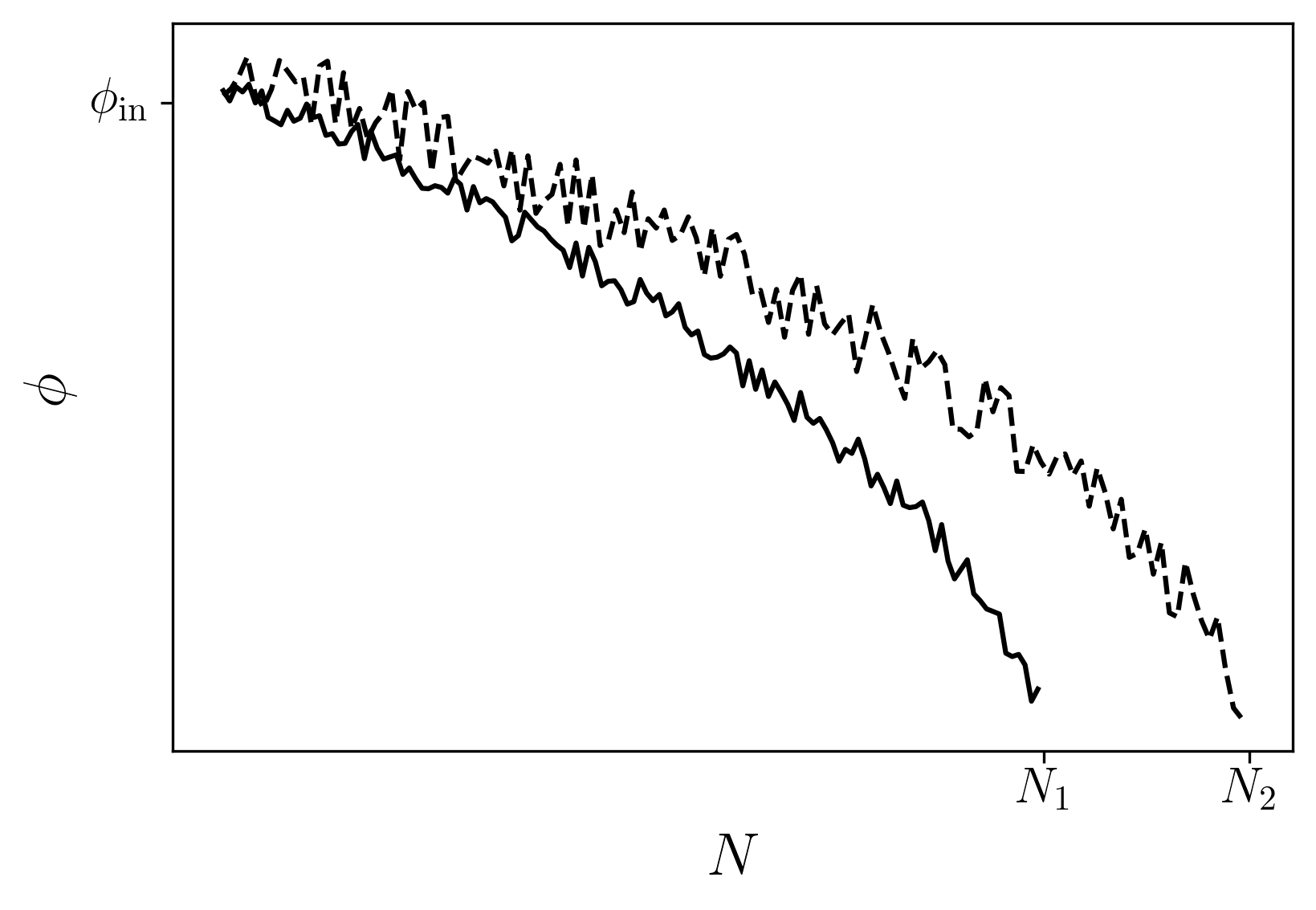}
\caption{An illustration showing the variation of the coarse-grained inflaton field during two different realizations. We have not given any units because this figure is only for illustrative purposes. The fluctuations have been amplified to make them more visible.}
\label{fig:trajectories}
\end{figure}
Typically, the term $\langle \zeta(\bm{x})^{2} \rangle$ would imply an ensemble average over many realizations of a particular size of the Universe in question. The meaning of the angular brackets in the context of stochastic inflation is as follows. Since the inflaton evolution is governed by SDEs, the coarse-grained trajectory will deviate from the background one (See Fig.\ref{fig:trajectories}). Over a large number of realizations these deviations away from the background enables one to compute the statistics of the fields. We denote $\phi_{\text{bg}}$ as the background value of the inflaton (without the noise terms)\cite{Ezquiaga:2018gbw,Levasseur:2014ska,Kunze:2006tu}. Hence, if the SDEs are solved enough times, the $\langle \delta\phi^{2} \rangle$ in Eq. \eqref{eq:P_zeta_def} can be interpreted as a stochastic average over all the realizations
\begin{equation}
\langle \delta\phi^{2}_{\text{st}} \rangle=\frac{1}{n_{\text{sim}}}\sum_{i=1}^{n_{\text{sim}}}(\bar{\phi}-\phi_{\text{bg}})^{2}_{i}
\end{equation}
The subscript `st' now stands for the fact that the correlation function has been computed from the solutions of the SDEs over a large number of realizations, labelled by $n_{\text{sim}}$. In the same way, higher order quantities like $\langle \delta\phi^{3}_{\text{st}} \rangle,\langle \delta\phi^{4}_{\text{st}} \rangle$ and cross terms like $\langle  \delta\phi_{\text{st}} \delta\pi_{\text{st}} \rangle$ can also be computed. This is illustrated in Fig. \ref{fig:trajectories}, where the inflaton starts out at $\phi_{\text{in}}$ in two realizations and follow different trajectories due to varying realizations of the noise and, in general, inflation ends with different $e$-foldings. This feature has led to the stochastic-$\delta N$ formalism for computing the curvature power spectrum \cite{Fujita:2013cna,Fujita:2014tja,Vennin:2015hra}.

\section{Numerical solution of coarse-grained inflaton field}\label{sec:numerical_solution_of_coarse_grained}
In this section we present the salient features that are involved in numerically simulating SDEs. The SDEs will be simulated on an event-by-event basis using a finite difference scheme. We will only focus on white noise in the current work. The stochastic term involves sampling the noise from a normal distribution function with zero mean and a variance given by the inverse grid-size. The noise amplitude is computed exactly by solving the Mukhanov-Sasaki equations at each time step.
\subsection{Stochastic calculus on a discrete lattice}

The two-point correlation of a function is given by
\begin{equation}
 \langle f(x)f(x') \rangle  = \displaystyle{\frac{\displaystyle{\sum_{\mathrm{all \: random \: events}}} f(x) f(x') \quad\quad  }{\mathrm{number \: of \: random \: events}}}
\end{equation}
In the continuous case, a white noise random function is defined as
\begin{equation}
 \langle f(x)f(x') \rangle = M(x) \delta(x-x') \text{ and } \langle f(x) \rangle = 0
\label{whiteDef}
\end{equation}
with $M(x)$ as the normalization factor and all higher-order cumulants are required to vanish.  We will fix $M(x) = 1$ here. The event-by-event distribution of $f$'s fluctuations is therefore a normal distribution with finite variance and zero mean. One can infer that, in a discrete case, we will have the following
\begin{equation} 
\langle f(x_i)f(x_{i'})\rangle = \frac{\delta_{ii'}}{\Delta x} 
\end{equation}
The normalization factor in the above equation has been set to unity. The $\delta_{ii'}/\Delta x$ becomes a Dirac $\delta$-function in the limit $\Delta x \to 0$. Therefore we sample the white noise function $f$ from a normal distribution of mean $0$ and standard deviation $1/\sqrt{\Delta x} $.
We use a random number generator for a large number of instances (e.g. $10^6 $) to overcome the statistical noise. Subsequently, we find the following relations for correlation functions on a discrete lattice. The details of these calculations can be found in the Ref. \cite{De:2020yyx}.

\begin{align}
\langle f(x_i)f'(x_{i'})\rangle &= \frac{\delta_{i+1,i'}-\delta_{i,i'}}{\Delta x^2} \\
\langle f'(x_i)f'(x_{i'})\rangle &= -\frac{\delta_{i,i'+1}+\delta_{i,i'-1}-2\delta_{i,i'}}{\Delta x^3} \\
\left\langle \int_{x_i}^{x_1} f(x') dx' \int_{x_i}^{x_2} f(x') dx' \right\rangle &= \min(x_1,x_2) - x_i
\end{align}

\subsection{Modeling the noise terms}\label{sec:noise_model}
We begin with arguably the most important aspect of stochastic inflation, that of accurately modeling noise. We have come across the expression which encodes the two-point correlation functions of the fields and their cross-correlations in Eq. \eqref{eq:noise_correlation}. The noise correlation functions of the fields at equal spatial points are
\begin{align*}
\langle \xi_{\phi}(N_{1})\xi_{\phi}(N_{2}) \rangle&=\frac{k_{\sigma}^{3}(N_{1})}{2\pi^2}(1-\epsilon_{1}(N_{1}))|\delta\phi_{k=k_{\sigma}(N_{1})}|^{2}\delta(N_{1}-N_{2})\\
\langle\xi_{\pi}(N_{1})\xi_{\pi}(N_{2})\rangle&=\frac{k_{\sigma}^{3}(N_{1})}{2\pi^2}(1-\epsilon_{1}(N_{1}))|\delta\pi_{k=k_{\sigma}(N_{1})}|^{2}\delta(N_{1}-N_{2}) \numberthis
\end{align*} 
We recall that the coarse-grained dynamics is written as a pair of coupled SDEs in Eq. \eqref{eq:coarse_grain_evol}. Once we have the amplitude of the noise, we can source it from a normal distribution. Schematically,
\begin{equation}
\xi \sim \left(\text{Noise amplitude}\right)^{1/2} \times \mathcal{N}\left(0,\frac{1}{\sqrt{\Delta N}}\right)
\end{equation}
where $\mathcal{N}(\mu,\sigma^{2})$ denotes a normal distribution with mean $\mu$ and variance $\sigma^{2}$.  \\
\indent The SDEs are discretized using the Euler-Maruyama method \cite{higham,Strauss:2017ekl,BURRAGE2000171}
\begin{align*}\label{eq:coarse_grain_discrete}
\bar{\phi}_{i+1}&=\bar{\phi}_{i}+\bar{\pi}_{\phi,i}\Delta N + \xi_{\phi,i}\Delta N\\
\bar{\pi}_{\phi,i+1}&=\bar{\pi}_{\phi,i}-(3-\epsilon_{1,i})\left( \bar{\pi}_{\phi,i}+\frac{\partial_{\phi}V}{V}\bigg\lvert_{i} \right)\Delta N+\xi_{\pi,i}\Delta N \numberthis
\end{align*}
where the subscripts $i$ refer to each time step $N_{i}$ and $\Delta N = N_{i+1}-N_{i}$. From the discretized equations we see that, at each time step, the noise terms modify the inflaton trajectory. We can implement this using the slow-roll approximation. However, for an accurate calculation, we need to forgo such simplifications and model the noise exactly. It can be shown that the noise terms become
\begin{align*}\label{eq:noise_discrete}
\xi_{\phi}(N_{i})&=\left[\frac{k_{\sigma}^{3}(N_{i})}{2\pi^2}(1-\epsilon_{1}(N_{i}))|\delta\phi_{k=k_{\sigma(N_{i})}}|^{2}\right]^{1/2}\frac{\mathcal{N}(0,1)}{\sqrt{\Delta N}}\\
\xi_{\pi}(N_{i})&=\left[\frac{k_{\sigma}^{3}(N_{i})}{2\pi^2}(1-\epsilon_{1}(N_{i}))|\delta\pi_{k=k_{\sigma(N_{i})}}|^{2}\right]^{1/2}\frac{\mathcal{N}(0,1)}{\sqrt{\Delta N}}\numberthis
\end{align*}
Here we used
\begin{equation}
\mathcal{N}\left( 0,\frac{1}{\sqrt{\Delta N}} \right)=\frac{1}{\sqrt{\Delta N}}\mathcal{N}(0,1)
\end{equation}
So, at each time step $N_{i}$, the Fourier modes $\delta\phi_{k}$ and $\delta\pi_{k}$ need to be evaluated for the wavenumbers $k=k_{\sigma}(N_{i})$. We note $\delta(N_{i}-N_{j})\approx \frac{1}{\Delta N}\delta_{ij}$. One way to reduce computational time is to take the noise amplitude calculation out of the SDE simulation part. Hence, for computational efficiency, we define the noise as follows
\begin{align*}
\xi_{\phi}(N_{i})&=\xi_{\phi}^{\text{amp}}(N_{i})\frac{\mathcal{N}(0,1)}{\sqrt{\Delta N}}\\
\xi_{\pi}(N_{i})&=\xi_{\pi}^{\text{amp}}(N_{i})\frac{\mathcal{N}(0,1)}{\sqrt{\Delta N}} \numberthis
\end{align*}
where the noise amplitudes $\xi_{\phi(\pi)}^{\text{amp}}$ can be read off from Eq. \eqref{eq:noise_discrete}. Solving the Mukhanov-Sasaki equation using finite difference method is a poor choice because of its slow convergence. This has been checked using a finite difference solver for the Mukhanov-Sasaki equation, where it was found that the $|\delta\phi_{k}|^{2}$ and $|\delta\pi_{k}|^{2}$ evolve to the correct superhorizon values only when the step size is at least $\Delta N=10^{-5}$. Using a Runge-Kutta solver is a more practical approach and we have chosen a second order Runge-Kutta (RK2) to numerically solve the Mukhanov-Sasaki equation for which a step size of $\Delta N=10^{-3}$ proved to be sufficient.
\subsection{Numerical solution scheme}\label{sec:numerical}
Having defined how to model the noise in Sec. (\ref{sec:noise_model}), we can describe our strategy to solve the SDEs and compute the power spectrum of curvature perturbations. As a prerequisite, one needs to consider an inflaton potential $V(\phi)$ with a suitable initial condition $\phi_{\text{in}}$ such that a reasonable number of $e$-foldings are generated. The following is an outline of the steps that we have followed.
\begin{itemize}
\item For a given inflaton potential, we find an initial field value $\phi_{\text{in}}$ that can generate a substantial period of inflation for the fiducial run. There is no general rule to how many $e$-folds it should be, but 60-70 $e$-folds suffices for most cases. The initial field field velocity or $(\pi_{\phi})_{\text{in}}$ is set using $-\partial{V}/V|_{\phi_{\text{in}}}$, which is accurate since the inflaton is at the attractor phase even for a USR potential.
\item With the inflaton potential and suitable initial conditions, the background evolution can be obtained by solving Eq. \eqref{eq:coarse_grain_discrete} without the $\xi$ terms. With this solution, all other relevant quantities are computed, which includes $a,H,\epsilon_{1}$ and $\epsilon_{2}$. With the given $\phi_{\text{in}}$, we first obtain a fiducial number of $e$-foldings $N$ which is used to set the observable scale $k_{\star}=\unit[0.05]{Mpc^{-1}}$. We set the observable scale 10 $e$-folds in our fiducial run. This describes the number of $e$-folds of observable inflation. Consequently, this is also used to set the scale factor $a=a_{0}e^{N}$ to the correct value of $a_{0}$. In general, if $k_{\star}$ becomes superhorizon at $N_{\star}$ into the fiducial run, then
\begin{equation}
a_{0}=\frac{k_{\star}}{H(N_{\star})}e^{-N_{\star}}
\end{equation}
\item After the background evolution has been obtained, the Mukhanov-Sasaki equation is solved for all $k_{\sigma}(N)=\sigma a(N)H(N)$ starting from the observable scale. One needs to be mindful of the fact that, when the modes are evolved, their initial conditions are set deep inside the horizon when $k\ll aH$. This is precisely the reason why the observable scale is set a certain $e$-folds into the fiducial run. Since $\sigma \ll 1$, it needs to be ensured that enough background evolution is available for the mode evolutions to be calculated. This depends on how far in the super-Hubble regime the coarse-graining scale is set. Otherwise, if the observable scale was set at zero $e$-folds, then there would not have been any information on the background evolution to set the initial conditions for $\delta\phi_{k}$ and $\delta\pi_{k}$. For $\sigma=0.01$, setting the observable scale at $N_{\star}=10$ should contain enough information regarding the background evolution. The initial conditions are set using the Bunch-Davies vacuum
\begin{eqnarray}
&\Re&(\delta\phi_{k_{\sigma}})=\frac{1}{a\sqrt{2k_{\sigma}}}\;\;\;\;\;\Im(\delta\phi_{k_{\sigma}})=0 \nonumber\\
&\Re&(\delta\pi_{k_{\sigma}})=-\frac{1}{a\sqrt{2k_{\sigma}}}\;\;\,\Im(\delta\pi_{k_{\sigma}})=-\frac{k_{\sigma}}{aH}\frac{1}{a\sqrt{2k_{\sigma}}}
\end{eqnarray}
\item With the background evolution and noise amplitudes calculated, the SDEs can now be solved. Since noise is being added starting from the observable scale at $N_{\star}=10$, the appropriate initial conditions need to be set for $\bar{\phi}$ and $\bar{\pi}_{\phi}$. The noise is implemented in a straightforward manner as already explained. After the equations are solved, the quantity $\bar{\phi}-\phi_{\text{bg}}$ is calculated. The equations are looped over a large number of simulations with $n_{\text{sim}}\sim 10^{6}$ with which the different correlation functions are evaluated.
\item Finally the power spectrum of curvature perturbations is calculated from Eq. \eqref{eq:P_zeta_def}. Since we are working with the $e$-fold variable, the derivative should be taken with respect to $N$. This is accomplished in a straightforward manner by noting that
\begin{equation}
\frac{d\ln k}{dN}=\frac{da/dN}{a}+\frac{dH/dN}{H}=1-\epsilon_{1}
\end{equation}
Hence
\begin{equation}\label{eq:P_zeta_2}
\mathcal{P}_{\zeta}(N)=\frac{1}{1-\epsilon_{1}}\frac{d}{dN}\left( \frac{\langle\delta\phi^{2}_{\text{st}}\rangle}{2\epsilon_{1}} \right)
\end{equation}
\end{itemize}
We provide a schematic of the algorithm.\\
\newline
\begin{algorithm}[H]
\SetAlgoLined
Solve background evolution;\\
Set $\Delta N$, $a$, $H$, $\epsilon_{1}$ and $\epsilon_{2}$;\\
Set $\sigma=0.01$;\\
\For{ $N \in [N_{\text{initial}},N_{\text{final}}]$}{
	\hspace{1cm}$k_{N}=\sigma a_{N}H_{N}$;\\
    \hspace{1cm}Solve Mukhanov-Sasaki Eq. \eqref{eq:mukhanov_sasaki} for each $k_{N},N$;\\
    \hspace{1cm}\Return{$\delta\phi_{k_{N}},\delta\pi_{k_{N}}$};
}
Calculate $\Xi_{\phi \phi},\Xi_{\pi\pi}$ for each $N$ \\
\For{ $j \in \{1,2,3,\cdots, n_{\text{sim}}\}$}{
    \hspace{1cm}Generate random normal event $\mathcal{N}(0,1)$ for each $N$-step. \\
    \hspace{1cm}Multiply the random event with the amplitude from Eq. \eqref{eq:noise_discrete}\\
    \hspace{1cm}\For{each $N$-step}{
        \hspace{1cm} \hspace{1cm} Solve SDE corresponding to Eq. \eqref{eq:coarse_grain_discrete}
    }
}
\caption{Coarse-grained inflaton evolution algorithm}
\end{algorithm}

\section{Results from test potentials}\label{sec:potential_results}

In this section we will apply the numerical techniques described thus far to two test potentials. The first one is the simplest inflaton potential, that of chaotic inflation with a quadratic term in $\phi$. Inflation takes place completely in slow-roll and the numerical results from the exact modeling of the noise should reproduce the familiar slow-roll result. The second potential that will be studied is a modified Starobinsky-type potential. It has an inflection point which results in the USR phase where the curvature power spectrum is amplified. This potential is mainly studied to shine some light on the significant contributions of the noise terms during such a phase and possible implications for PBH formation.

\subsection{Chaotic potential}

We start with one of the simplest inflation potentials that can be constructed out of a single parameter. It is the quadratic potential or commonly called the chaotic inflation potential \cite{Linde:1983gd,Linde:1986fd}. It takes the following form

\begin{equation}
V(\phi)=\frac{1}{2}m^{2}\phi^{2}
\end{equation}
For this potential, we choose $N=64$ for the fiducial run and set the observable scale at $N_{\star}=10$ producing a total of $\Delta N=54$ $e$-folds of observable inflation. Using the pivot scale set at $N_{\star}$ where $k_{\star}=\unit[0.05]{Mpc^{-1}}$, the parameter $m^{2}$ is set to $4.42\times 10^{-11}M_{\text{pl}}^{2}$. Here we use the fact that $\mathcal{P}_{\zeta}$, under the slow-roll approximation at CMB scales, is given by

\begin{figure}
\centering
\includegraphics[scale=0.8]{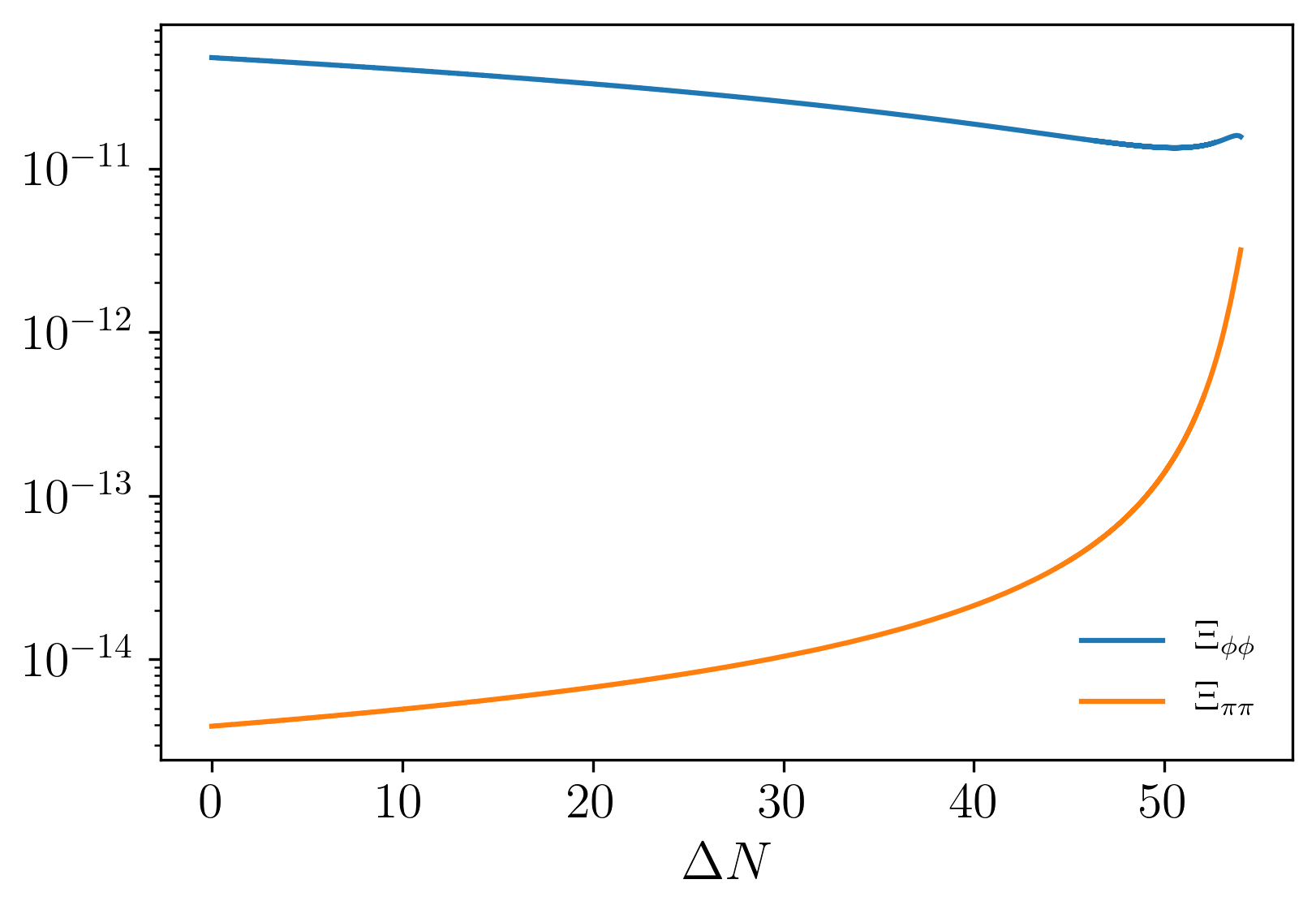}
\caption{Evolution of correlation functions $\Xi_{\phi\phi}$ and $\Xi_{\pi\pi}$ for $\sigma=0.01$. The $\Xi_{\pi\pi}$ noise is subdominant as already derived under the slow-roll approximation.}
\label{fig:chaotic_corr}
\end{figure}
\begin{figure}
\centering
\includegraphics[scale=0.8]{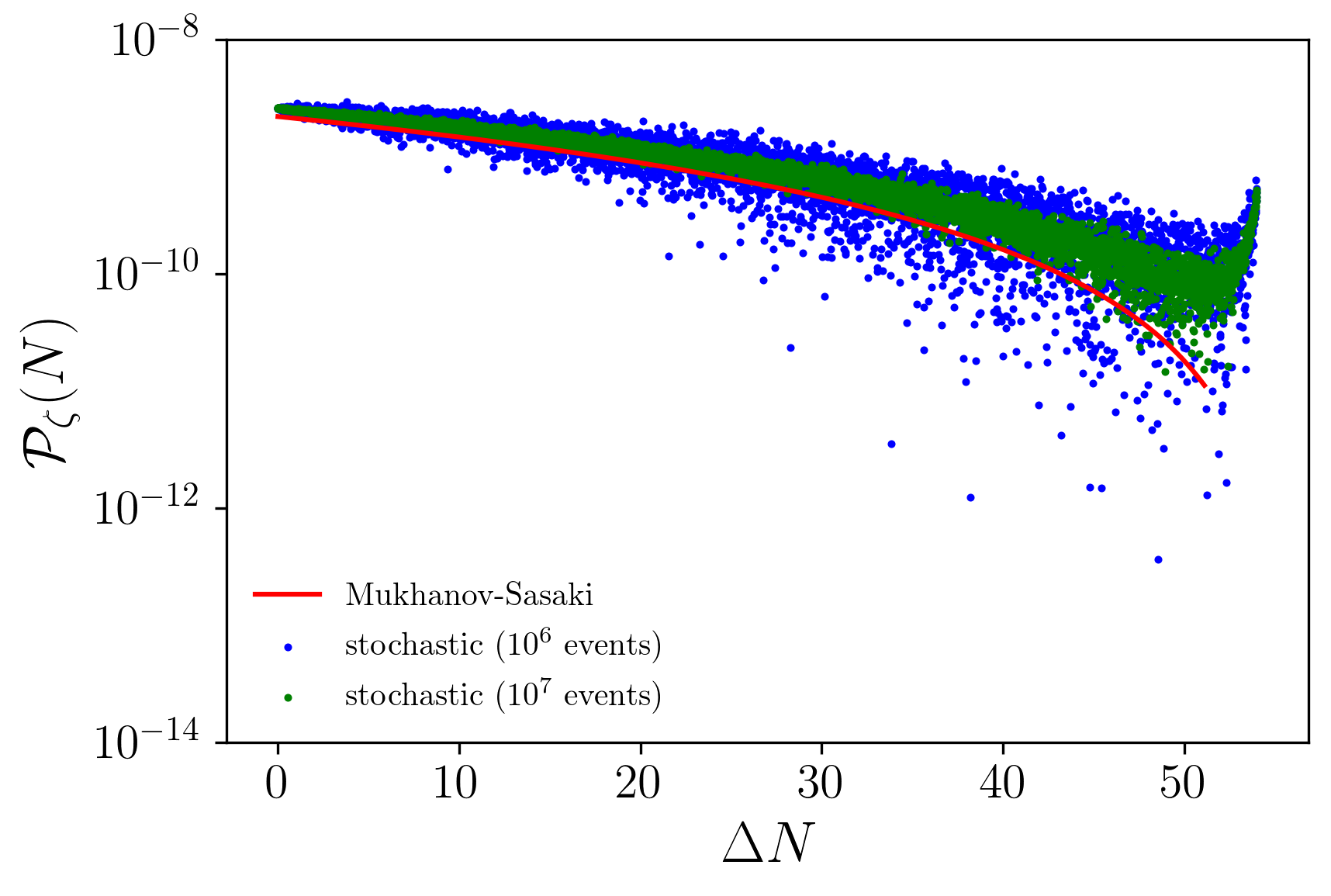}
\caption{Power spectrum of curvature perturbations for the chaotic potential using $\sigma=0.01$. The blue and green dotted curves represent results using Eq. \eqref{eq:P_zeta_2} for $10^6$ and $10^7$ realizations respectively while the solid red curve is the solution obtained from solving the Mukhanov-Sasaki equation.}
\label{fig:chaotic_power}
\end{figure}
\begin{figure}
\centering
\includegraphics[scale=0.8]{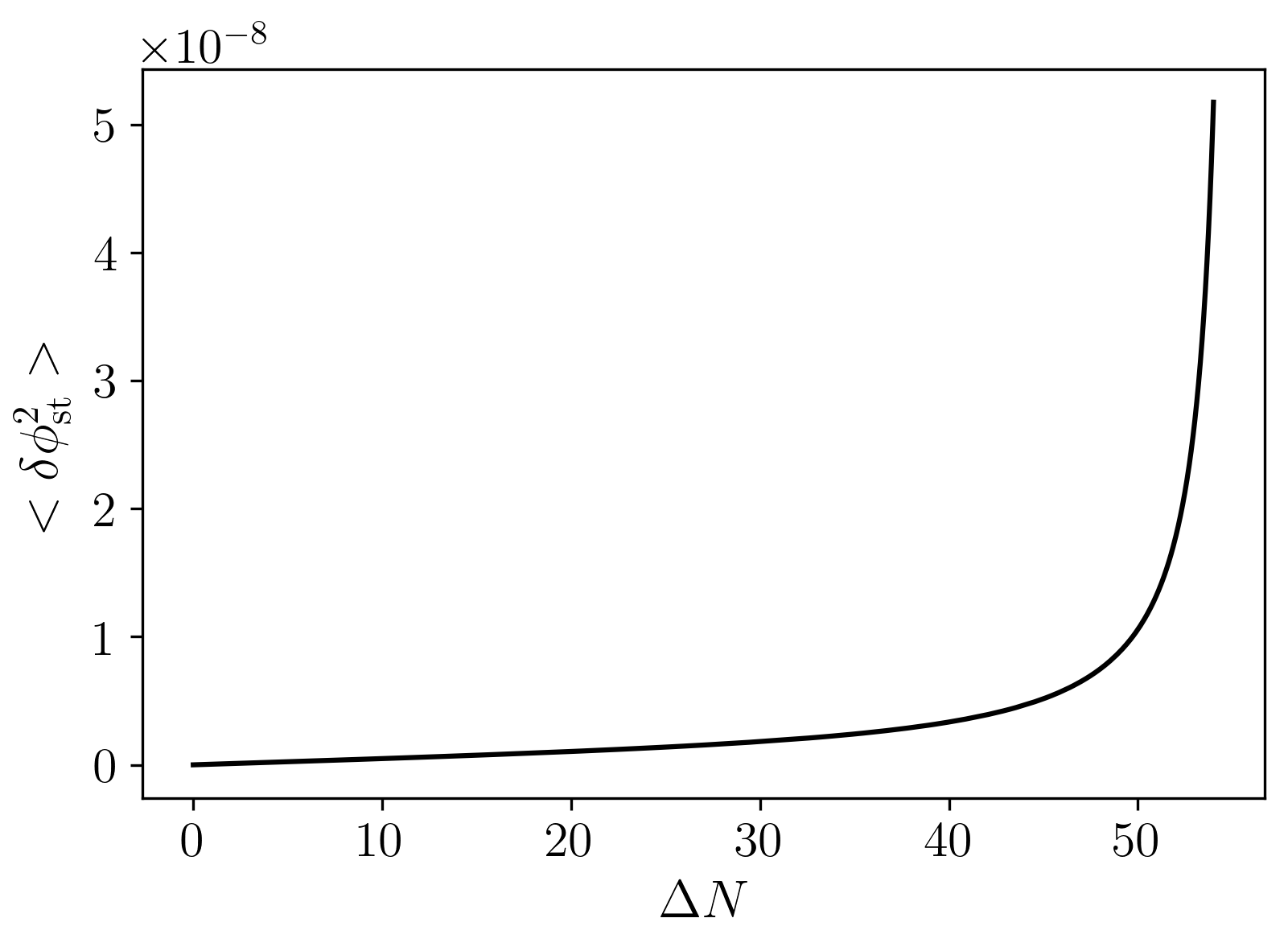}
\includegraphics[scale=0.8]{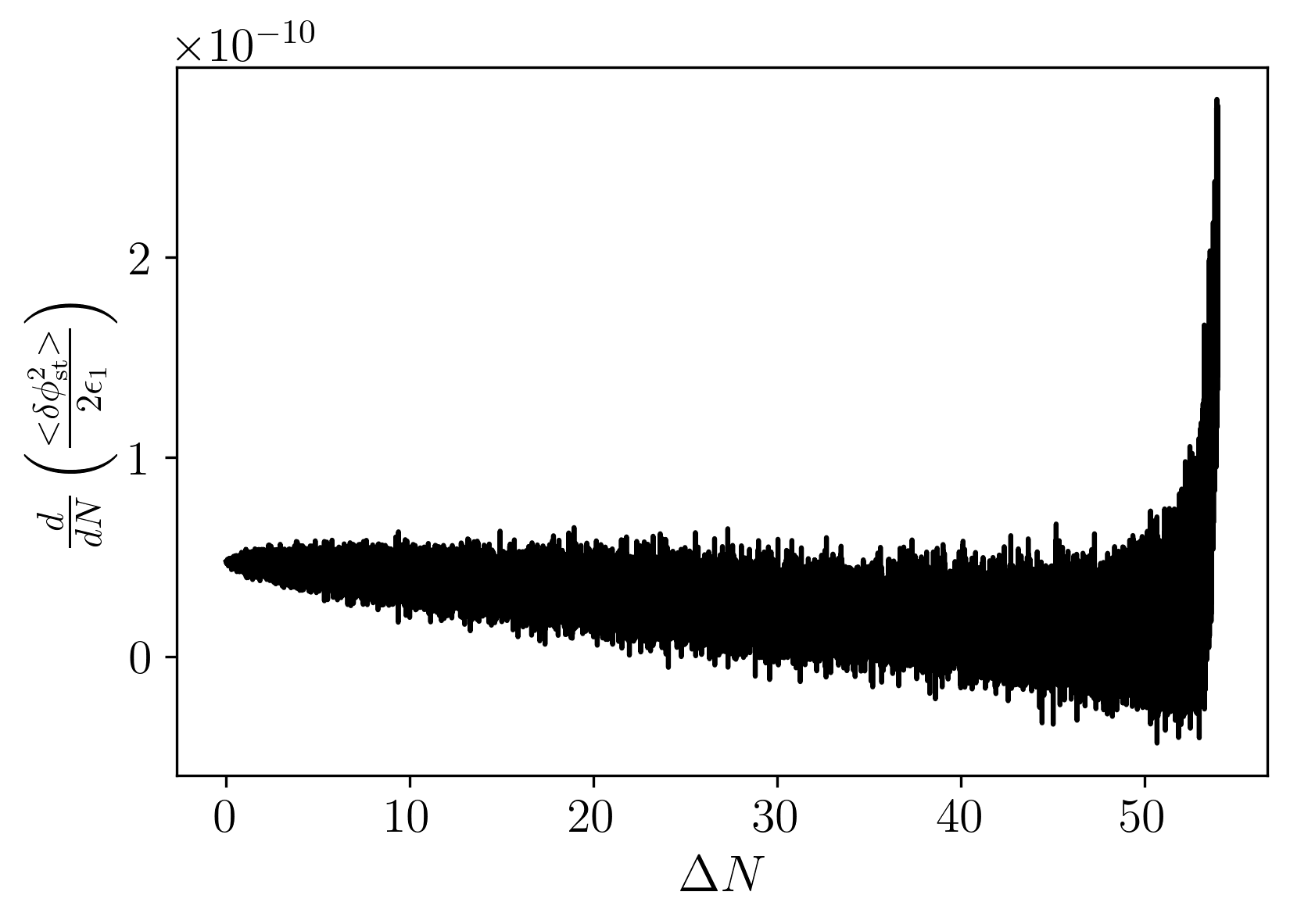}
\caption{Comparison of $\langle \delta\phi^{2}_{\text{st}} \rangle$ and its derivative. They are in units of $M_{\text{pl}}^{2}$. We see that while the former appears smooth, the latter is not.}
\label{fig:Adel}
\end{figure}


\begin{equation}
\mathcal{P}_{\zeta}(k_{\star})=\frac{H^{2}(k_{\star})}{8\pi^{2}\epsilon_{1}(k_{\star})}\simeq 2.2\times 10^{-9}
\end{equation}
We first check how the two correlation functions $\Xi_{\phi\phi}$ and $\Xi_{\pi\pi}$ evolve with time. These are illustrated in Fig. \ref{fig:chaotic_corr} for $\sigma=0.01$. It is seen that $\Xi_{\phi\phi}$ is the dominant contributor to the stochastic noise and $\Xi_{\pi\pi}$ is suppressed by a few orders of magnitude. At this point, we can also compute the curvature power spectrum $\mathcal{P}_{\zeta}$ using Eq. \eqref{eq:P_zeta_2} for a large number of realizations and compare with the result obtained by solving the Mukhanov-Sasaki equation. This is shown in Fig. \ref{fig:chaotic_power} for $10^6$ and $10^7$ realizations of the SDEs. At first glance, the curves look rather noisy. Although this can be rectified by increasing the number of simulation events, it might not be a feasible option from a computational standpoint. A major source of noisy randomness is the derivative of stochastic quantities. This can be seen in Fig. \ref{fig:Adel} where the figure in the right panel corresponds to $\frac{d}{dN} \langle \delta\phi^{2}_{\text{st}}/2\epsilon_{1} \rangle$ for $10^{6}$ realizations. \\
\indent One way to remove this derivative from the definition of the power spectrum is to define the moments of the fluctuations as
\begin{equation}\label{eq:field_moments}
\langle \delta\phi^{n}_{\text{st}}\delta\pi^{m}_{\text{st}} \rangle=\int d\bar{\phi}d\bar{\pi}(\bar{\phi}-\phi_{\text{bg}}(N))^{n}(\bar{\pi}_{\phi}-\pi_{\text{bg}}(N))^{m}P(\bar{\phi},\bar{\pi}_{\phi},N)
\end{equation}
where $P(\bar{\phi},\bar{\pi}_{\phi},N)=P(\bm{\Phi},N)$ is the probability distribution of the coarse-grained inflaton in phase-space defined by the Fokker-Planck equation \cite{Ezquiaga:2018gbw,Vennin:2015hra,1994PhRvD..50.6357S}

\begin{equation}\label{eq:fokker_planck}
\frac{dP(\bm{\Phi},N)}{dN}=-\frac{\partial}{\partial\Phi_{A}}\left( D_{A}P(\bm{\Phi},N)-\frac{\Xi_{AB}}{2}\frac{\partial P(\bm{\Phi},N)}{\partial\Phi_{B}} \right)
\end{equation}
The subscripts $A,B=1,2$ run over the coarse-grained inflation field and its conjugate momentum. The term $D_{A}$ refer to the drift components defined by
\begin{align*}
D_{\phi}&=\bar{\pi}_{\phi}\\
D_{\pi}&=-(3-\epsilon_{1})\left( \bar{\pi}_{\phi}+\frac{\partial_{\phi}V}{V} \right) \numberthis
\end{align*}

\begin{figure}
\centering
\includegraphics[scale=0.8]{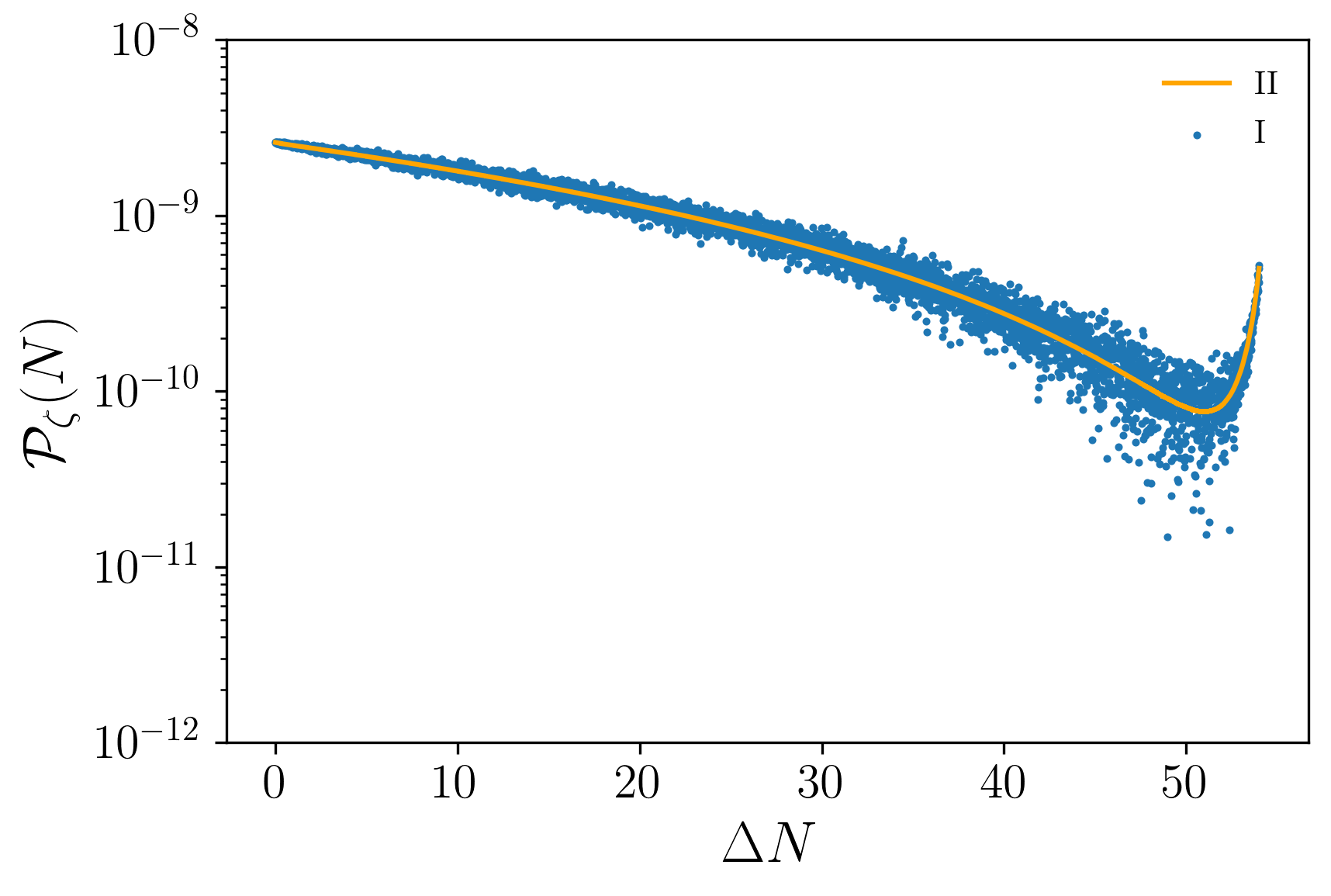}
\caption{Comparison of $\mathcal{P}_{\zeta}$ obtained from Eqs. \eqref{eq:P_zeta_2} and \eqref{eq:P_zeta_approx} labelled as I and II respectively. The orange curve passes through the points and has no noisy features.}
\label{fig:chaotic_power4}
\end{figure}

These can be used to construct an expression for $\mathcal{P}_{\zeta}$ that does not contain any derivatives with respect to $N$. We first note that the term in the parenthesis in Eq. \eqref{eq:P_zeta_2} can be expanded into 

\begin{equation}
\frac{d}{dN}\left(\frac{\langle \delta\phi^{2}_{\text{st}} \rangle}{2\epsilon_{1}}\right)=\frac{1}{2\epsilon_{1}}\left( \frac{d}{dN}\langle \delta\phi^{2}_{\text{st}} \rangle-\epsilon_{2}\langle \delta\phi^{2}_{\text{st}} \rangle \right)
\end{equation}
We refer here to Appendix \ref{sec:correlation_FP} of the paper (or Appendix B of Ref. \cite{Ezquiaga:2018gbw}) where the time evolution of any $n$-point correlation function is derived using Fokker-Planck equation. The derivative of $\langle \delta\phi_{\text{st}}^{2} \rangle$ can be expressed as

\begin{equation}
\frac{d}{dN}\langle \delta\phi^{2}_{\text{st}} \rangle=\Xi_{\phi\phi}+2\langle \delta\phi_{\text{st}}\delta\pi_{\text{st}} \rangle
\end{equation}
such that the power spectrum can be expressed in the form

\begin{equation}\label{eq:P_zeta_approx}
\mathcal{P}_{\zeta}=\frac{1}{1-\epsilon_{1}}\frac{1}{2\epsilon_{1}}\left(\Xi_{\phi\phi}+2\langle \delta\phi_{\text{st}}\delta\pi_{\text{st}} \rangle-\epsilon_{2}\langle \delta\phi^{2}_{\text{st}} \rangle \right)
\end{equation}
With this, the noisy behaviour of the power spectrum can be overcome and $\mathcal{P}_{\zeta}$ becomes a smooth curve. However, one should be aware of the fact that the prescription for eliminating the derivatives of the correlation function is an approximation. As explained in \cite{Ezquiaga:2018gbw}, to completely specify $P(\bm{\Phi},N)$, one would need to solve an infinite system of first order differential equations. However, the series can be truncated at the desired order at which all higher order moments are subdominant. Refering  to Appendix \ref{sec:correlation_FP}, in Eq. \eqref{eq:correlation_derivative}, we get an expansion in the derivative of $\ln V$. For the derivative of a correlation function of the form $\langle \delta\phi_{\text{st}}^{n}\delta\pi_{\text{st}}^{m} \rangle$, the series can be truncated at the order $(n+m)$. In our case, $n=2$ and $m=0$. In Fig. \ref{fig:chaotic_power4}, we see the curvature power spectrums obtained from Eq. \eqref{eq:P_zeta_2} and from Eq. \eqref{eq:P_zeta_approx}. The solid curve obtained using Eq. \eqref{eq:P_zeta_approx} displays no noisy features for $10^6$ realizations whereas the other curve is quite noisy. 

\subsection{Deformed Starobinsky potential}\label{sec:deformed_Starobinsky}

\begin{figure}
\centering
\includegraphics[scale=0.8]{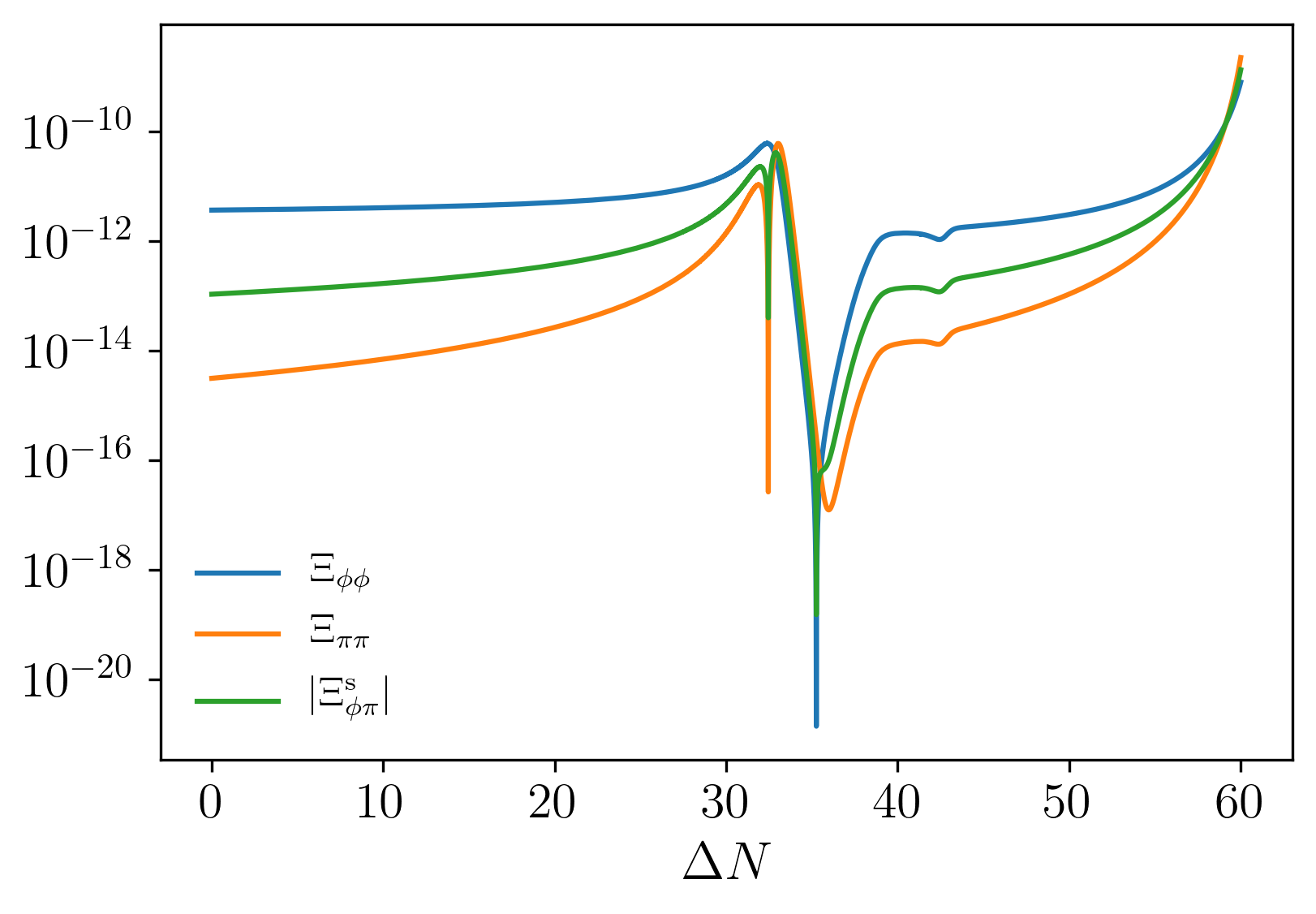}
\includegraphics[scale=0.8]{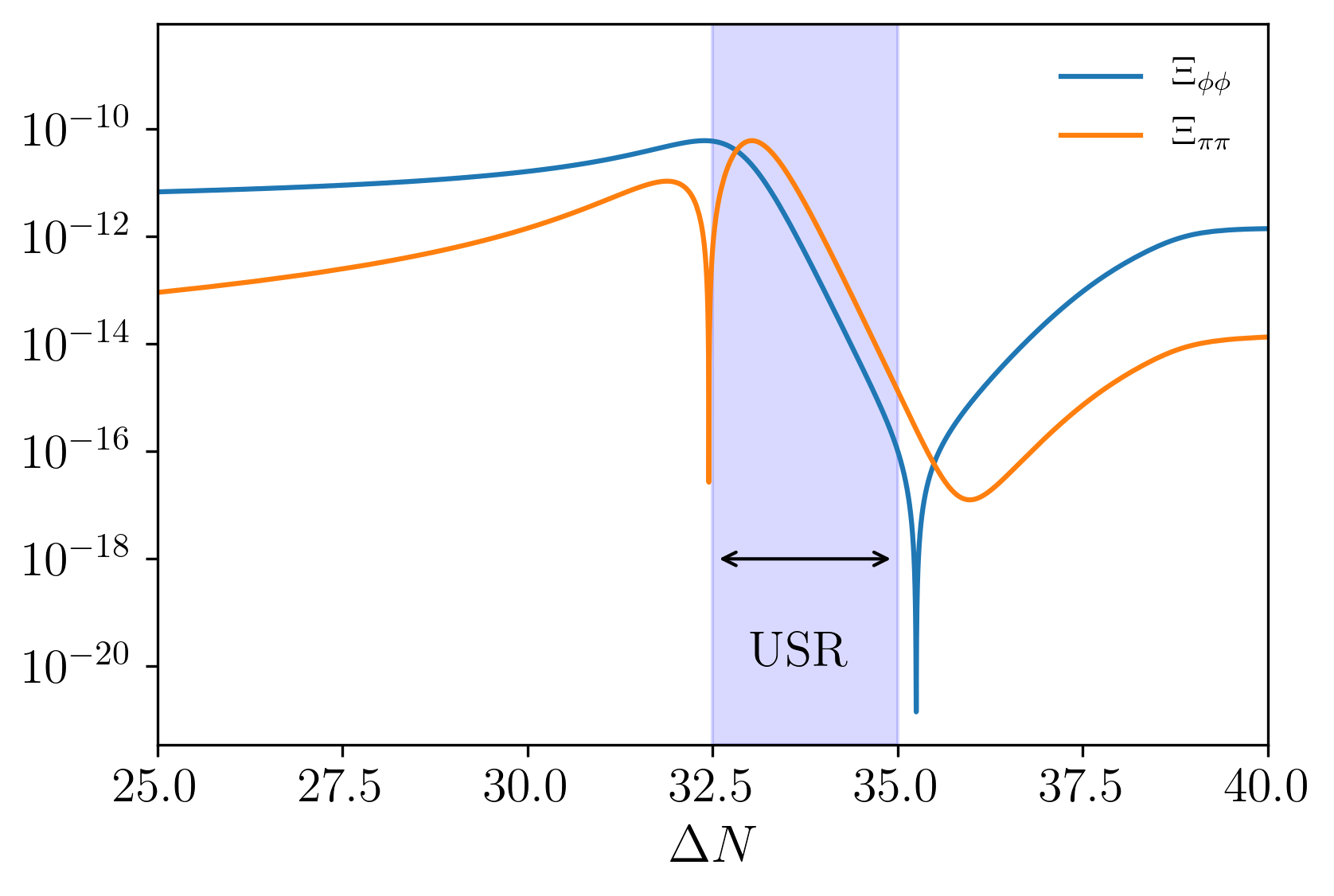}
\caption{Evolution of the correlation functions $\Xi_{\phi\phi},\Xi_{\pi\pi}$ and $\Xi^{\text{s}}_{\phi\pi}$ for $\sigma=0.01$. It is clear that the $\pi-\pi$ noise becomes significant in the USR phase.}
\label{fig:deformed_starobinsky_correlations}
\end{figure}

\begin{figure}
\centering
\includegraphics[scale=0.8]{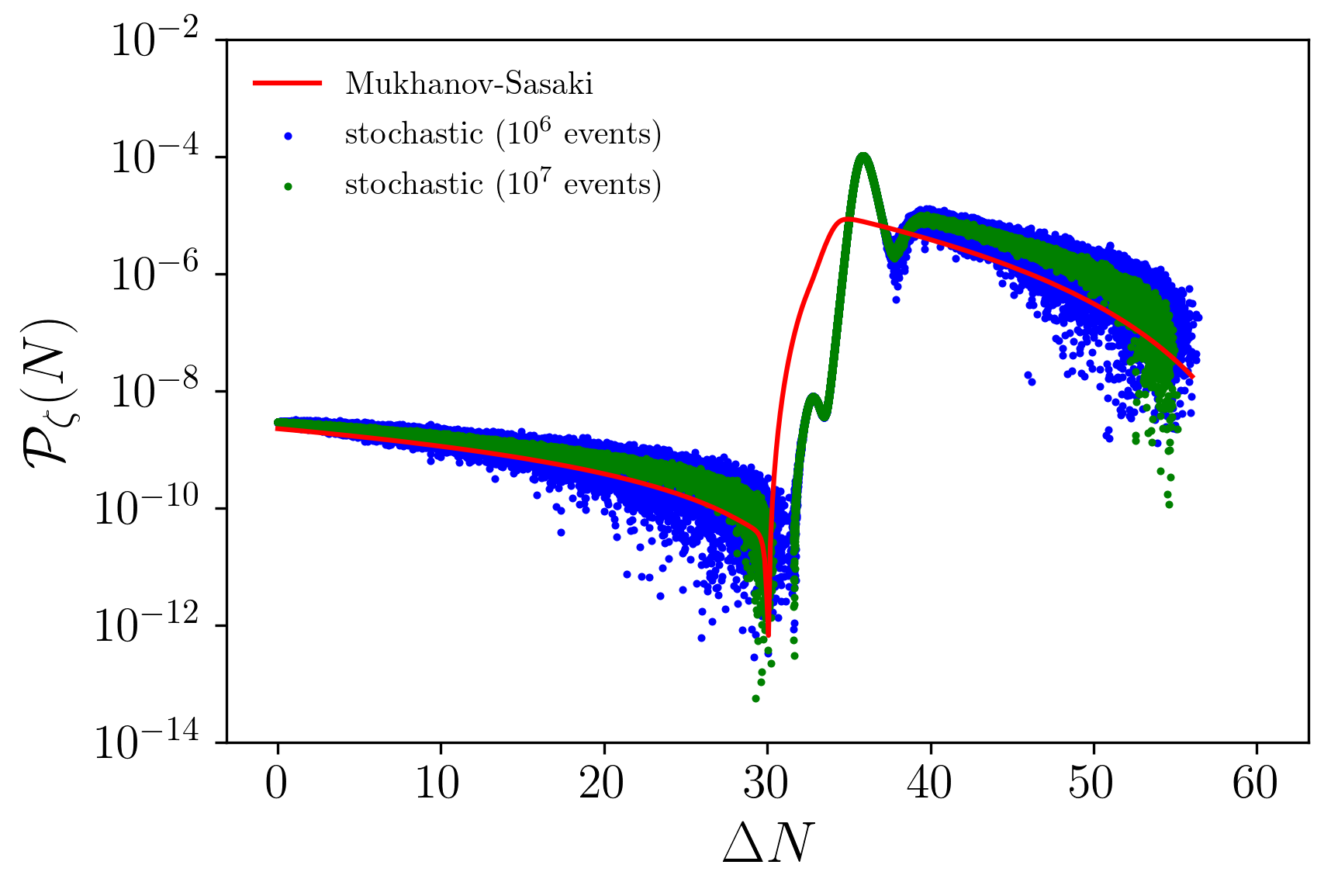}
\caption{Power spectrum of curvature perturbations for the deformed Starobinsky potential using $\sigma=0.01$. The blue and green dotted curves represent results using Eq. \eqref{eq:P_zeta_2} for $10^6$ and $10^7$ realizations respectively while the solid red curve is the solution obtained from solving the Mukhanov-Sasaki equation.}
\label{fig:deformed_power}
\end{figure}

The numerical tools that have been developed so far can be used to study a class of USR inflation models. Such inflation models typically possess some peculiar features in their potentials that create departures from slow-roll behaviour. The presence of an inflection point in the potential can slow down the inflaton and give rise to amplifications in the curvature power spectrum. This type of potential was discussed in \cite{Ketov:2018uel} as a ways of modifying the familiar Starobinsky inflation in the context of supergravity. We consider a potential of the following form \cite{PhysRevD.101.023533}

\begin{equation}
V(\phi)=V_{0}\left( 1+\xi-e^{-\alpha\phi}-\xi e^{-\beta\phi^{2}} \right)^{2}
\end{equation}
In the limit $\xi\to 0$, the potential reduces to the $R+R^{2}$ modification of Einstein gravity, which gives rise to Starobinsky inflation. A potential of this form is of interest in cases where PBH production is concerned. However, one should be wary of the fact that USR models are generally very fine-tuned.\footnote{Readers are referred to \cite{Hertzberg:2017dkh} for a detailed and quantitative study of the level of fine-tuning required for PBH formation using a polynomial potential with an inflection point.} There are technically three free parameters in this potential since $V_{0}$ is fixed by the CMB normalization of the power spectrum. The parameter $\alpha$ can be set to $\sqrt{2/3}$ for it to retain similarity with the Starobinsky model. The remaining parameters are then adjusted depending on the type of inflaton dynamics that is required. We impose the condition that there is an inflection point of $V(\phi)$ at $\tilde{\phi}$. This leads to

\begin{align*}\label{eq:inflection_condition}
\xi&=-\frac{\alpha}{2\beta\tilde{\phi}}e^{-\alpha\tilde{\phi}+\beta\tilde{\phi}^{2}}\\
\beta&=\frac{\beta\tilde{\phi}+1}{2\tilde{\phi}^{2}}\numberthis
\end{align*}
\begin{figure}
\centering
\includegraphics[scale=0.8]{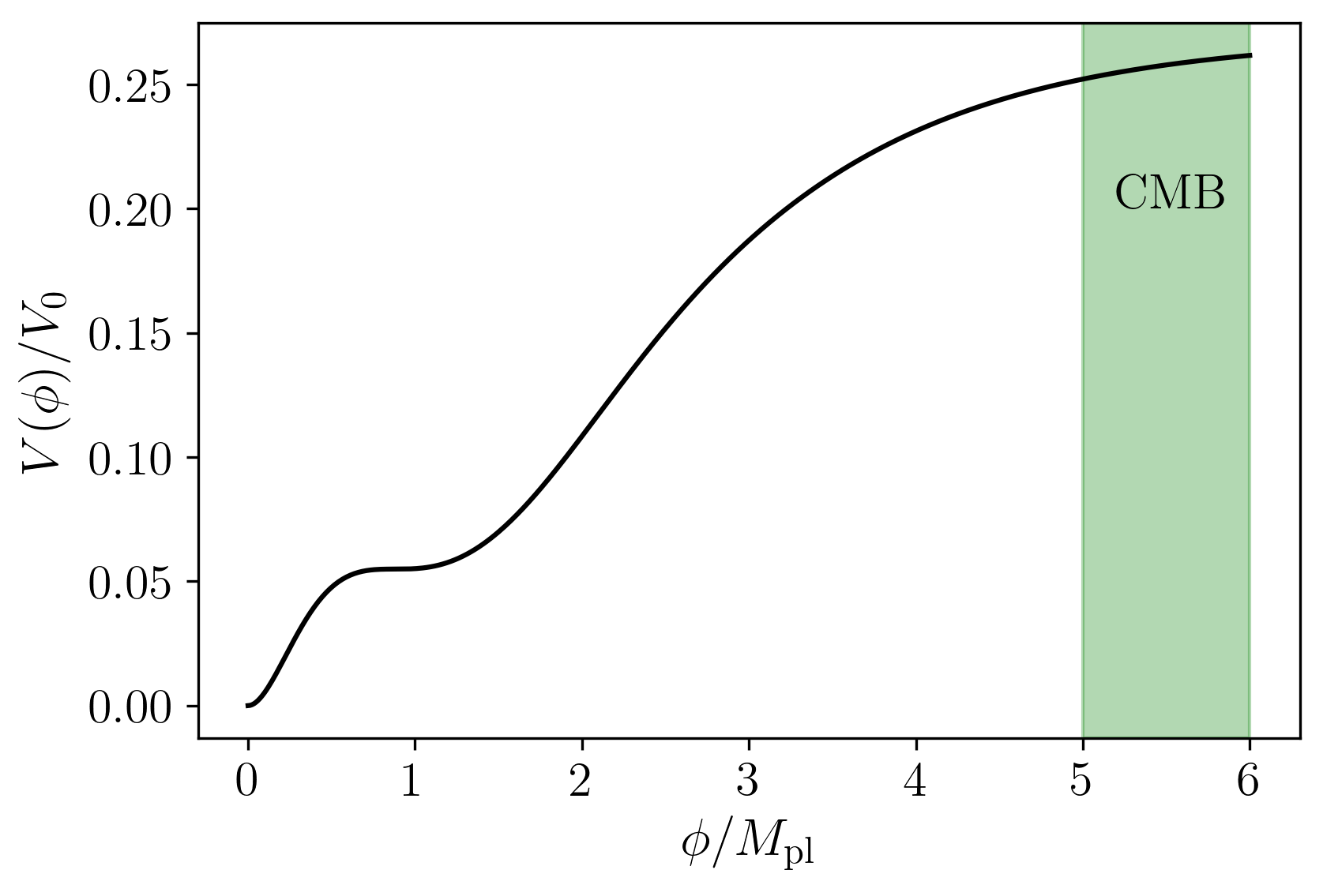}
\includegraphics[scale=0.8]{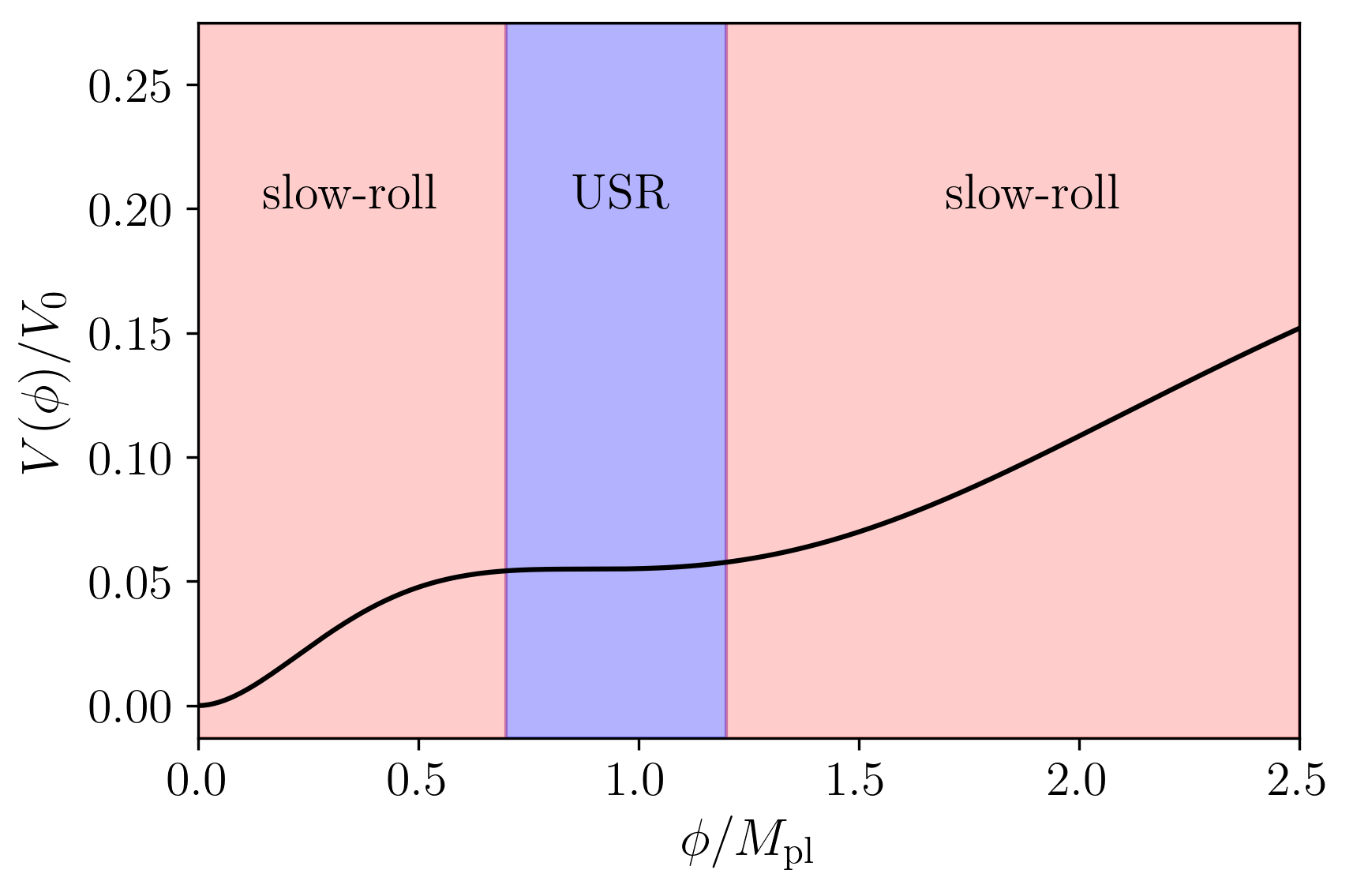}
\caption{The deformed Starobinsky potential for super-Planckian field excursions. The shaded regions show the different periods in the inflationary stage.}
\label{fig:potentials}
\end{figure}
Using Eq. \eqref{eq:inflection_condition}, the parameters can be suitably selected to yield the desired background evolution. The potential is shown in Fig. \ref{fig:potentials}. In the top subfigure, the inflation starts out in the shaded region. These are the comoving scales which later reenter the horizon during the formation of the CMB. The inflaton evolution takes place under slow-roll for the most part, except for the shaded blue region in the right panel which is the non-attractor USR period. We chose the following parameter set: $V_{0}=1.27\times 10^{-9}M_{\text{pl}}^{4}$, $\alpha=\sqrt{2/3}$, $\beta =1.114905$ and $\xi=-0.480964$. For the numerical simulations, we set $\phi_{\text{in}}=5.82M_{\text{pl}}$ which produces approximately $N=70$ $e$-folds of inflation. Like the chaotic potential case, we fix the observable scale at $N_{\star}=10$ where we start adding the noise terms to the SDEs.\\
\indent In Fig. \ref{fig:deformed_starobinsky_correlations} we plot the the noise correlation functions as a function of $e$-folds. During the early stages of the inflation, much like slow-roll, the $\Xi_{\pi\pi}$ term is subdominant. Once the inflation enters the USR phase, $\Xi_{\pi\pi}$ becomes comparable to $\Xi_{\phi\phi}$ and can no longer be ignored. As a result, one should expect significant difference between the behavior of background fields and noise-incorporated fields. We already know that curvature power spectrum is enhanced near an inflection point\cite{PhysRevD.101.023533}. We can see this semi-quantitatively in the following way: near an inflection point $\partial_{\phi}V\simeq\partial_{\phi\phi}V=0$ and the inflaton evolution simplifies to

\begin{equation}
\frac{d^{2}\phi}{dN^{2}}+(3-\epsilon_{1})\frac{d\phi}{dN}\simeq 0
\end{equation}
the solution of which can be expressed as 
\begin{equation}
\phi(N)\sim\int^N e^{-\int^{N'} (3-\epsilon_{1})dN''}dN'
\end{equation}
Then the curvature power spectrum behaves in the following way near the inflection point
\begin{align*}
\mathcal{P}_{\zeta}^{1/2}&=\frac{H^{2}}{2\pi\dot{\phi}}
=\frac{H}{2\pi d\phi/dN}\sim \frac{H}{2\pi}\left[ \int e^{-\int (3-\epsilon_{1})dN''}dN' \right]^{-1}\numberthis
\end{align*}
As long as $\epsilon_{1}<3$, there is an exponential amplification of the curvature power spectrum near the vicinity of the inflection point. If we disregard $\epsilon_{1}$ for a moment and consider that the USR phase lasts for $\delta N$ $e$-folds, the power spectrum scales as $\mathcal{P}_{\zeta}\sim e^{6\delta N}$. Now we can compare the standard result of $\mathcal{P}_{\zeta}$ computed by solving the Mukhanov-Sasaki equation with that of the stochastic procedure. The results are plotted in Fig. \ref{fig:deformed_power} where the blue and green dotted curves represent the stochastic result for $10^{6}$ and $10^{7}$ realizations of the SDEs respectively. As is evident, there is an $\mathcal{O}(1)$ enhancement in $\mathcal{P}_{\zeta}$ relative to the Mukhanov-Sasaki result. The peak occurs at $\Delta N_{\text{peak}} = 35.9$ which, in terms of the comoving wavenumber, is around $k_{\text{peak}}\sim \unit[8.63\times 10^{13}]{Mpc^{-1}}$. Concerning PBH formation, comoving scales of this size would collapse to form PBHs of mass close to $\unit[6.6\times 10^{17}]{g}$. The peak in the curvature power spectrum is approximately $\mathcal{P}_{\zeta}^{\text{peak}}\simeq 10^{-4}$. Although this is not nearly large enough to collapse to produce PBHs in sufficient abundances, it indicates that there are parameter sets which can work in favor of PBH formation. Due to the added amplification in the power spectrum, less finely tuned parameter sets can be used to explain PBH formation. `Fine-tuning' here refers to finding the subspace of parameters in the entire parameter space that would yield desired results. A major issue regarding inflection-point potentials is that the mere presence of an inflection point does not guarantee that inflation ends in a finite number of $e$-folds. In fact, for a fairly vast number of parameters, the inflaton gets stuck in the plateau and $\epsilon_{1}=1$ is never reached. Depending on the frequency of parameter sets, such USR models can be described among the following cases.

\begin{itemize}
\item Case I: Inflation does not end in a finite number of $e$-folds. (many parameter sets)
\item Case II: Inflation ends in a finite number of $e$-folds and $\mathcal{P}_{\zeta}^{\text{max}}\sim 10^{-6}-10^{-5}$. (a few parameter sets)
\item Case III: Inflation ends in a finite number of $e$-folds and $\mathcal{P}_{\zeta}^{\text{max}}\sim 10^{-3}-10^{-2}$. (very few parameter sets)
\end{itemize}
Hence, in cases such as this, fine-tuning can refer to a distinction between cases II and III. The added enhancement in $\mathcal{P}_{\zeta}$ would help explore PBH formation for parameter sets which follow case II.\\
\indent The deformed Starobinsky potential is not just a toy model as it can be constructed using supergravity models with chiral superfields $\Phi$ and $S$ with $\Phi$ having shift symmetry. Such models predict an F-term scalar potential of the form \cite{Kawasaki:2000yn,Ketov:2014qha,Kallosh:2010xz}
\begin{equation}
V=f^{2}(\chi)
\end{equation}
where $\chi\sim \Im\Phi$ and $f$ is an arbitrary holonomic function that enters via the superpotential.

\section{Discussion and comparison with slow-roll noise}

In Sec. \ref{sec:potential_results}, we computed the power spectrum of curvature perturbations by simulating the SDEs over millions of realizations for quadratic and deformed Starobinsky potentials. In this section, we discuss the results and compare them with those using slow-roll noise. Although the parameters in both potentials have been chosen such that $\mathcal{P}_{\zeta}\sim 2.2\times 10^{-9}$ at the pivot scale $k_{\star}=\unit[0.05]{Mpc^{-1}}$ \cite{Akrami:2018odb}, the stochastic results predict slighly larger values. In our computations, we chose $N_{\star}=10$ into the fiducial run as corresponding to $k_{\star}$. The results obtained in these computations were

\begin{itemize}
\item Chaotic: $\mathcal{P}_{\zeta}(k_{\star})=2.61\times 10^{-9}$
\item Deformed Starobinsky: $\mathcal{P}_{\zeta}(k_{\star})=2.95\times 10^{-9}$
\end{itemize}
These values are larger than the observable obtained from CMB measurements. However, we see that these discrepancies are not present when the slow-roll expressions for the noise terms are used. As stated in a previous section, under the slow-roll approximation the noise terms take the following forms

\begin{align*}
\langle \xi_{\phi}(N_{1})\xi_{\phi}(N_{2}) \rangle&\simeq\frac{H^{2}}{4\pi^{2}}\delta(N_{1}-N_{2})\\
\langle \xi_{\pi}(N_{1})\xi_{\pi}(N_{2}) \rangle&\simeq 0 \numberthis
\end{align*}

\begin{figure}[t]
\centering
\includegraphics[scale=0.8]{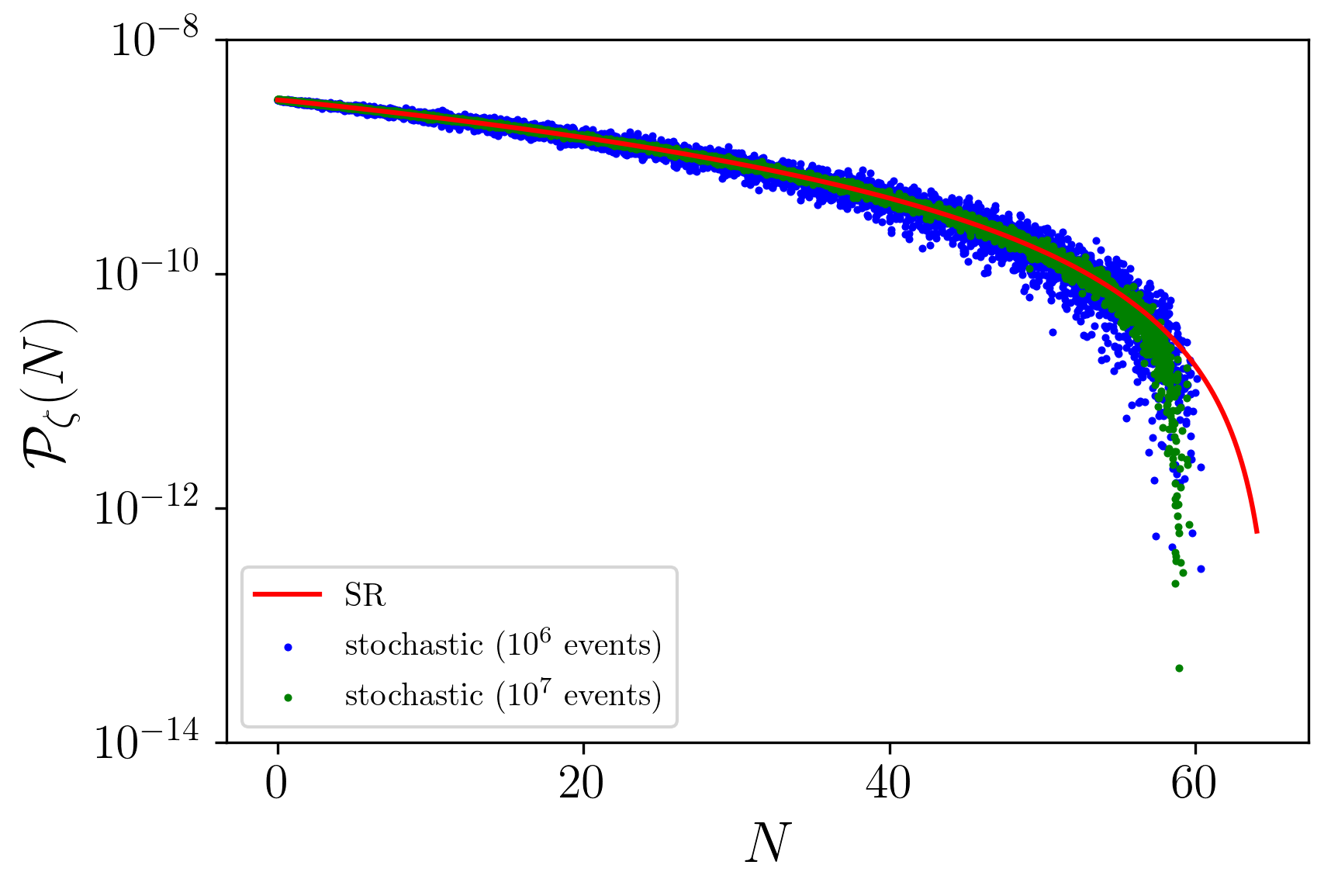}
\includegraphics[scale=0.8]{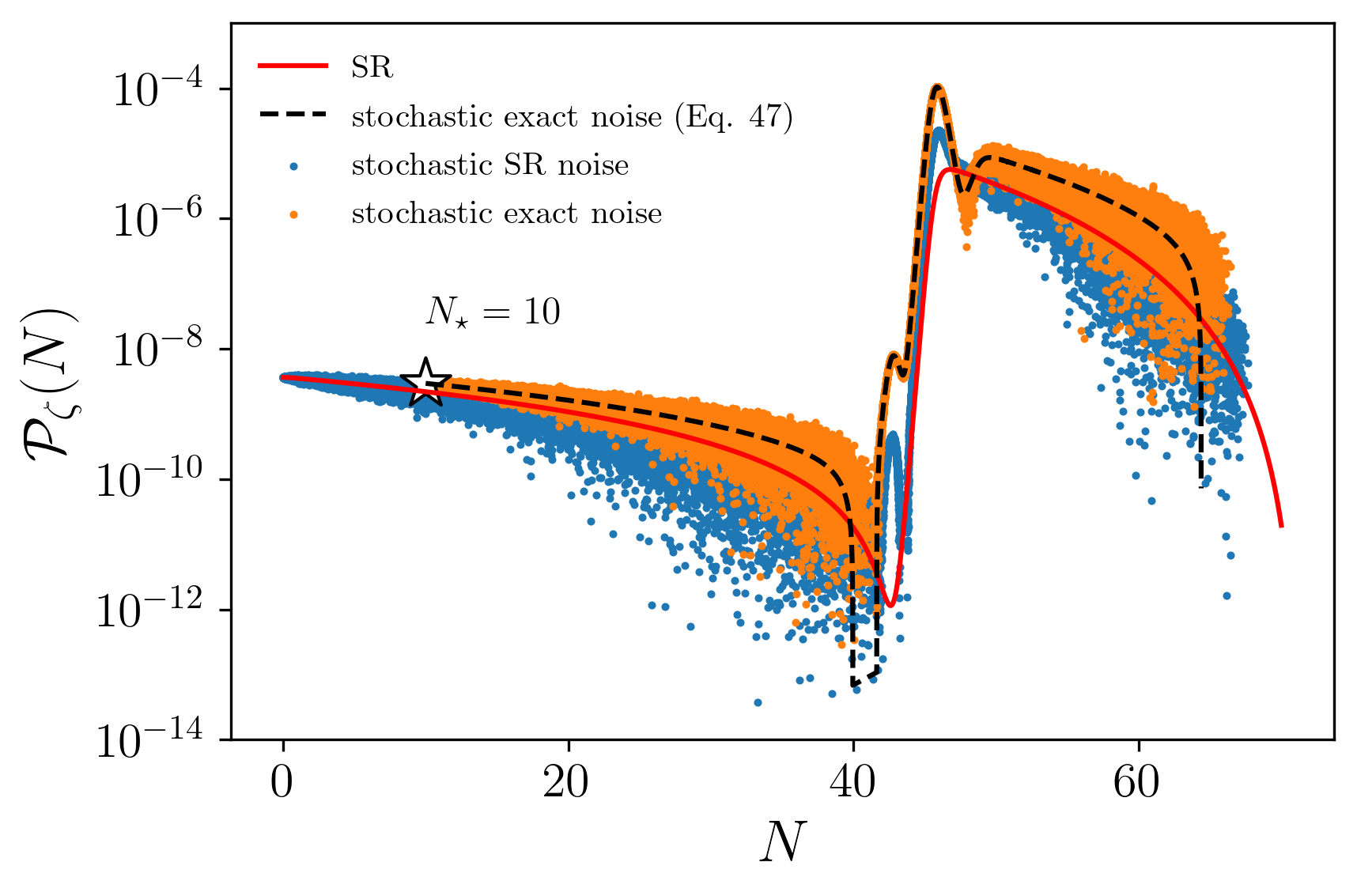}
\caption{Power spectrum of curvature perturbations computed using slow-roll noise from the chaotic (top panel) and deformed Starobinsky potentials (bottom panel) respectively. The power spectrum has been computed for $10^{6}$ realizations of the stochastic process using $\sigma=0.01$. The bottom panel contains both slow-roll and exact noise, along with the data obtained from Eq. \eqref{eq:P_zeta_approx}.}
\label{fig:P_zeta_SR}
\end{figure}

In Fig. \ref{fig:P_zeta_SR} we show the plots for $\mathcal{P}_{\zeta}$ for the chaotic and deformed Starobinsky potentials calculated using slow-roll noise.\footnote{We should note that, in Fig. \ref{fig:P_zeta_SR} the $N$ refer to total number of $e$-folds and not the observable number. This is reasonable since the Mukhanov-Sasaki equation does not have to be solved and the noise can be added to each time trivially.} The solid red line in the plots represent the slow-roll result which is computed using $\mathcal{P}_{\zeta}^{\text{SR}}=H^{2}/8\pi^{2}\epsilon_{1}$. The $\mathcal{P}_{\zeta}$ at 10 $e$-folds does produce the desired CMB normalization and the deformed Starobinsky slow-roll power spectrum is reproduced rather well. For the deformed Starobinsky potential, we also plot the power spectrum computed using the exact noise, for which the data points start from $N_{\star}=10$ which is the chosen observable scale.

\begin{figure}
\centering
\includegraphics[scale=0.8]{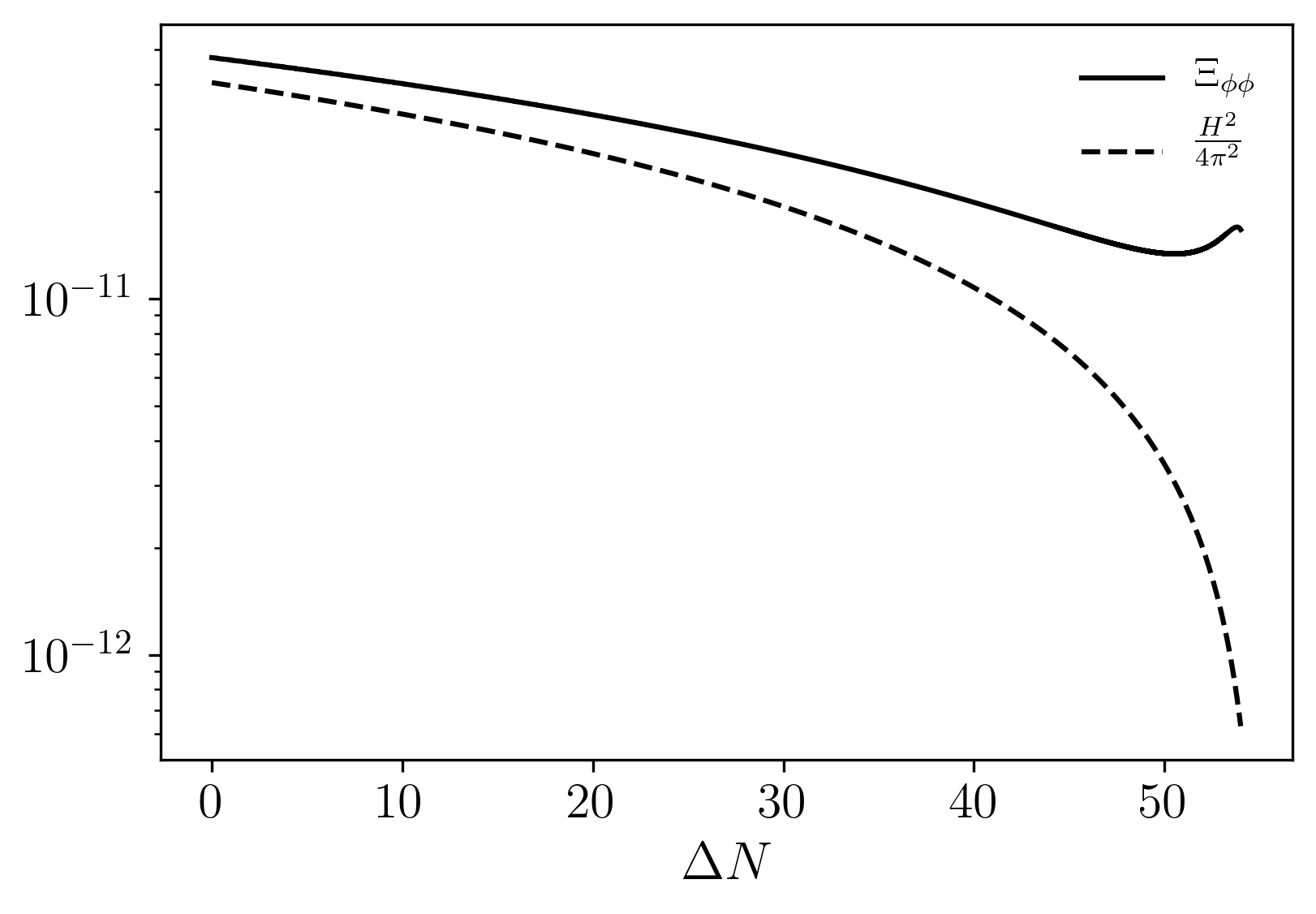}
\caption{Comparison of the exact $\phi-\phi$ noise (solid black) with the slow-roll approximation (dotted black).}
\label{fig:chaotic_noise}
\end{figure}

Let us discuss the discrepancy that occurs when the unapproximated noise is used. The reason is the fact that, even for complete slow-roll, the $\Xi_{\phi\phi}$ noise does not exactly correspond to $H^{2}/4\pi^{2}$. In Fig. \ref{fig:chaotic_noise}, we plot the exact numerical calculation of the noise along with $H^{2}/4\pi^{2}$. Therefore, at each time step in the SDEs, the amount of noise being added is slightly different than $H^{2}/4\pi^{2}$, which is then reflected in the calculation of $\mathcal{P}_{\zeta}$.  

It is also interesting to note that the additional amplification in $\mathcal{P}_{\zeta}$ near the inflection point is not as large as shown in \cite{Ezquiaga:2018gbw}. There, compared to the Mukhanov-Sasaki solution of the power spectrum, the stochastic result showed a $\mathcal{O}(10^{3})$ increase. In our case, the amplification turned out to be only of order $\mathcal{O}(10)$. 

\begin{figure}
\centering
\includegraphics[scale=0.8]{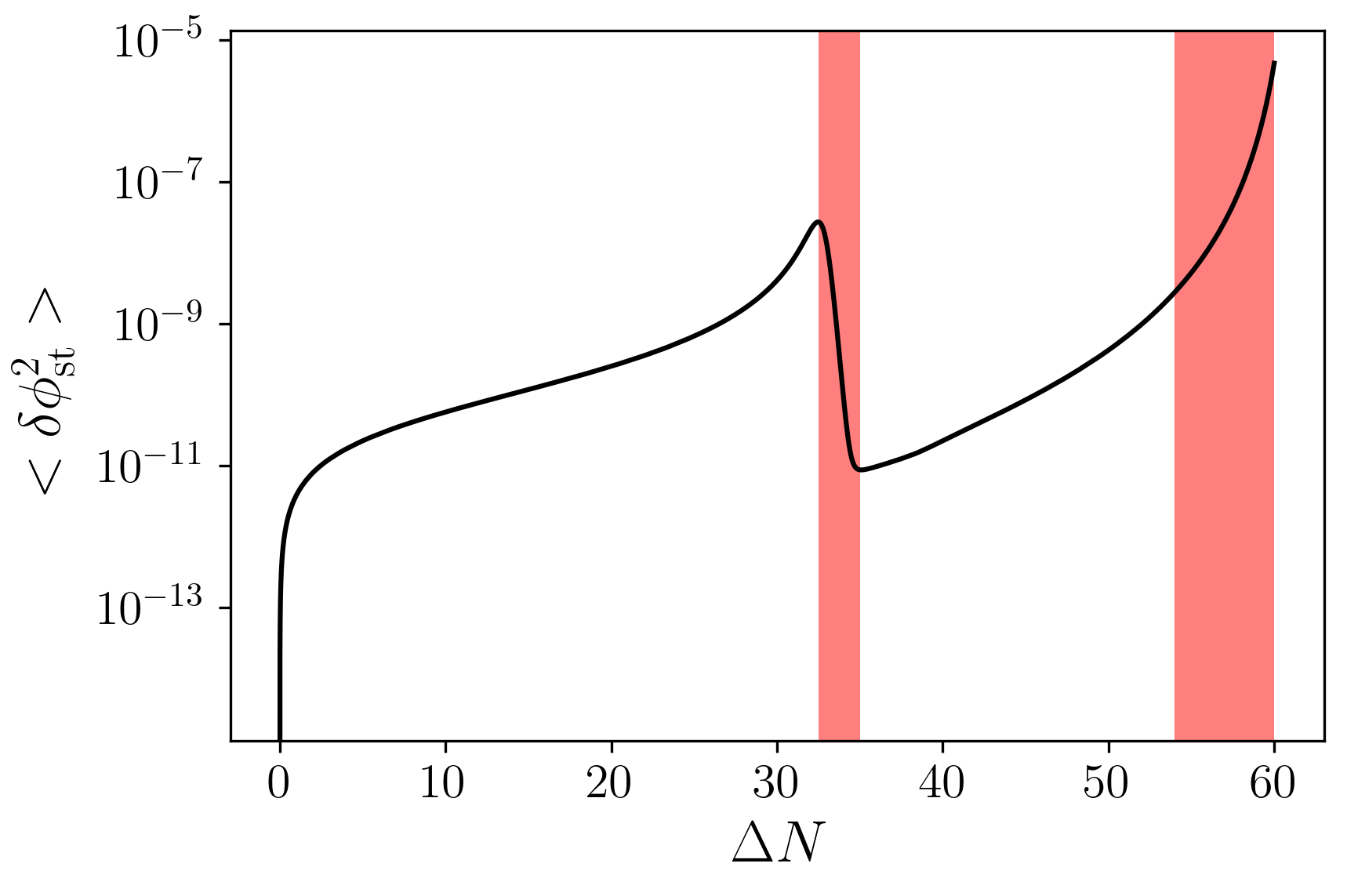}
\caption{The plot of $\langle \delta\phi_{\text{st}}^{2} \rangle$ for the deformed Starobinsky potential for $\sigma=0.01$. The two shaded regions indicated the regions where $\mathcal{P}_{\zeta}$ become negative. In the left shaded region $\langle \delta\phi_{\text{st}}^{2} \rangle$ decreases and hence has a negative slope.}
\label{fig:AdeltaPhi2_Deformed_Starobinsky}
\end{figure}

Finally, we comment on the missing points in $\mathcal{P}_{\zeta}$ of the deformed Starobinsky potential, as seen in Fig. \ref{fig:deformed_power}. At around $\Delta N\sim 30$ $e$-folds, the values of $\mathcal{P}_{\zeta}$ become negative and therefore do not show up on a logarithmically scaled $y$-axis. The reason for this occuring lies in the form of $\langle \delta\phi_{\text{st}}^{2} \rangle$. We recall that, $\mathcal{P}_{\zeta}\sim\frac{d}{dN}\langle \delta\phi_{\text{st}}^{2} \rangle -\epsilon_{2}\langle \delta\phi_{\text{st}}^{2} \rangle$. In Fig. \ref{fig:AdeltaPhi2_Deformed_Starobinsky} we show two regions, shaded in red, responsible for $\mathcal{P}_{\zeta}$ acquiring negative values. The one on the left is where $\langle \delta\phi_{\text{st}}^{2} \rangle$ decreases with $e$-fold and, as a result, the derivative acquires negative values. The one on the right, for $\Delta N>54$ is where $-\epsilon_{2}\langle \delta\phi_{\text{st}}^{2} \rangle$ term becomes larger than $\frac{d\langle \delta\phi_{\text{st}}^{2} \rangle}{dN}$, which also results in negative values.\\

\begin{figure}
\centering
\includegraphics[scale=0.8]{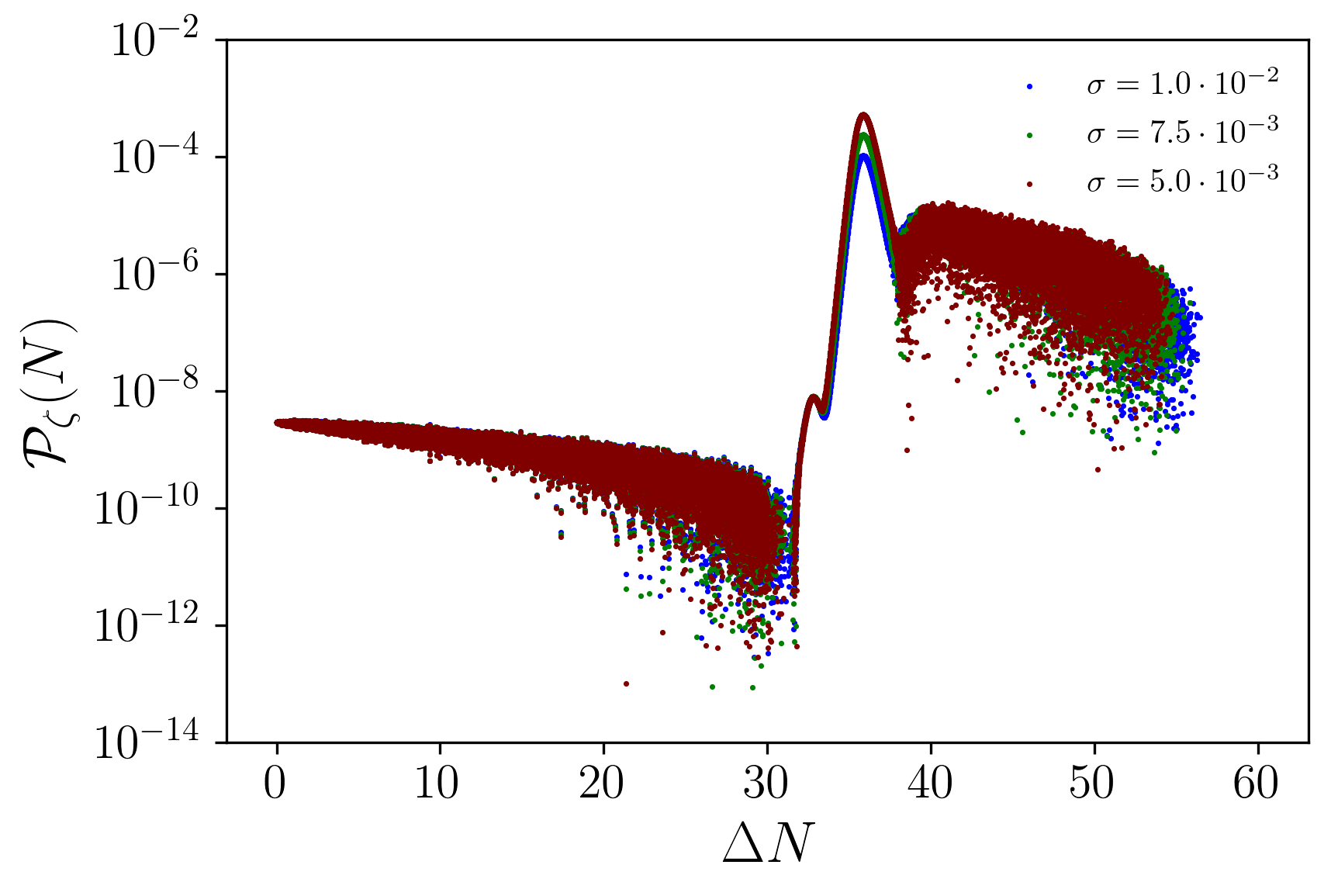}
\includegraphics[scale=0.8]{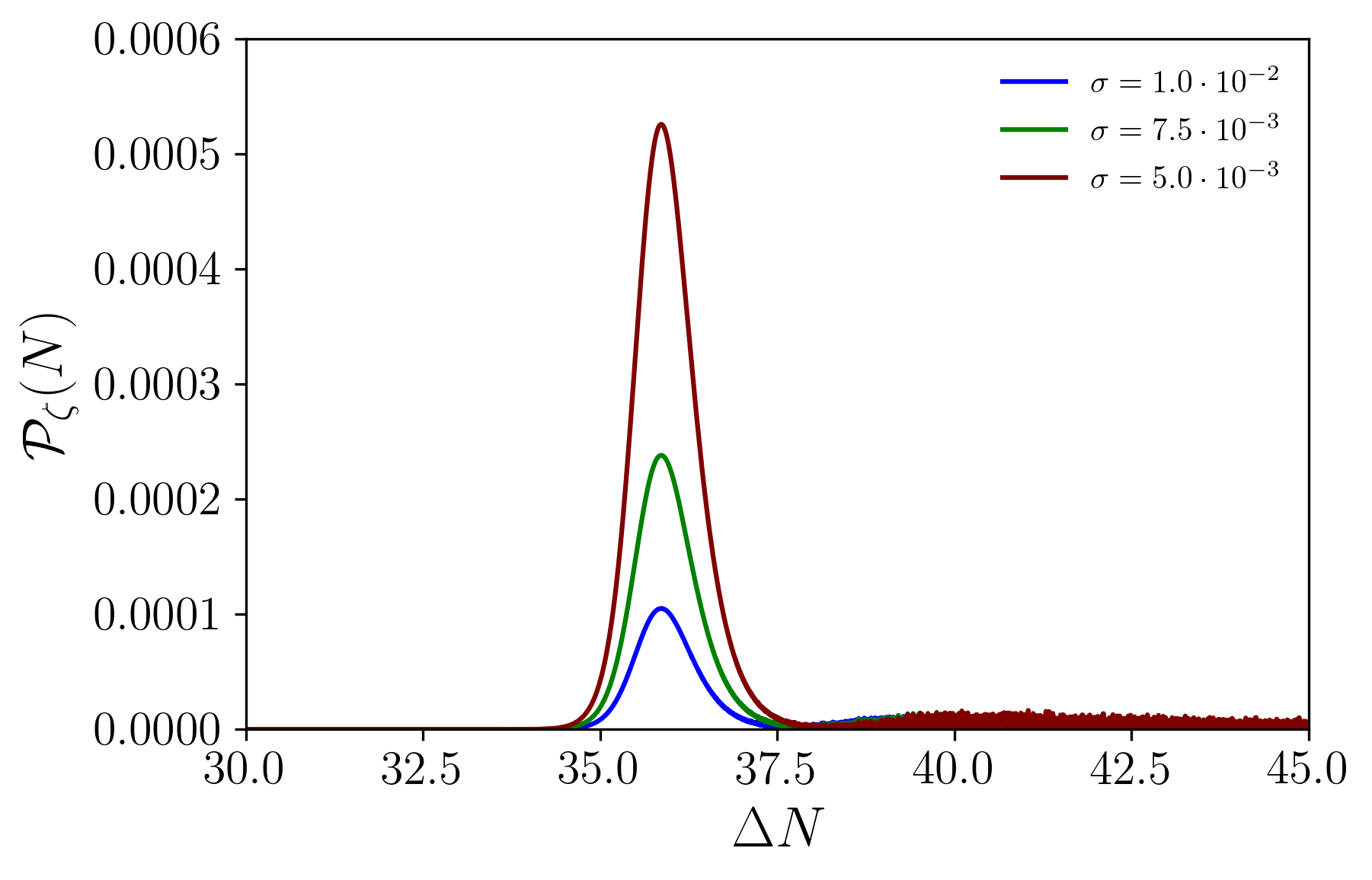}
\caption{Power spectrum of curvature perturbations for the deformed Starobinsky potential using $\sigma=1.0\times 10^{-2}$ (blue), $\sigma=7.5\times 10^{-3}$ (green) and $\sigma=5.0\times 10^{-3}$ (magenta). The bottom panel is a magnified version of the peaks in the top panel. There is a relative increase in the size of the peak going from the first to last.}
\label{fig:P_zeta_different_sigma}
\end{figure}

 It should also be checked whether the choice of $\sigma$ affects the result of $\mathcal{P}_{\zeta}$. A different choice of $\sigma$ does not produce any changes for the quadratic potential, as can be expected. But it is a slightly different story with the deformed starobinsky potential. We carry out computations for the same parameter set with $N=70$ for $\sigma =5\times 10^{-3}$. Coincidentally, for this value of $\sigma$, the $k_{\sigma}$ would correspond to the smallest wavenumber for which the evolution can be numerically computed since, for anything smaller, there would not be enough background evolution information. We plot the $P_{\zeta}$ in Fig. \ref{fig:P_zeta_different_sigma} for $\sigma=10^{-2},7.5\times 10^{-3}$ and $5\times 10^{-3}$ for $10^{6}$ realizations of the SDEs. We observe that, although the shape of $\mathcal{P}_{\zeta}$ stays similar, there is an increase in the size of the peak, the largest of which is of the order $\mathcal{P}_{\zeta}^{\text{max}}\sim 5\times 10^{-4}$ for $\sigma=5\times 10^{-3}$. Ideally, this could also be tested for a smaller $\sigma$. However setting $N_{\star}=10$ effectively admits $k_{\sigma}\approx 5\times 10^{-3}aH$ as the smallest coarse-graining scale for which the evolution can be computed. If PBH formation is under investigation, then the $\mathcal{P}_{\zeta}$ obtained from $\sigma=5\times 10^{-4}$ might be sufficient. But it is also known that PBH abundance depends on the choice of window functions \cite{Ando:2018qdb,Young:2019osy} and whether one uses peaks theory or Press-Schechter \cite{Yoo:2018kvb,Mahbub:2020row}. This investigation is deferred to a future work.

\section{Jackknife error estimation}\label{sec:jackknife}
One of the common ways to reduce bias and estimate errors in stochastic modeling is the jackknife method \cite{Efron,Efron2}. This method involves resampling of the data, specifically `sampling without replacement'. It estimates the error of statistics without making any assumptions about the distribution that generated the data. We create jackknife samples by sequentially deleting a single observation from the sample, or in other words, creating ``leave-one-out" data sets. In our case, we consider the two-point correlation statistic $S$ on the original sample size of $10^6$ events. We leave out the $i_{\text{th}}$ event to create the $i_{\text{th}}$ jackknife statistic $S_i$. The average of the jackknife samples is $S_{\rm avg} = \sum_i S_i/n$. The jackknife error is then estimated as 
\begin{equation}
\sigma_{\rm jack} = \sqrt{\frac{n-1}{n} \sum_i (S_i-S_{\rm avg})^2}
\end{equation}

\begin{figure}
\centering
\includegraphics[scale=0.8]{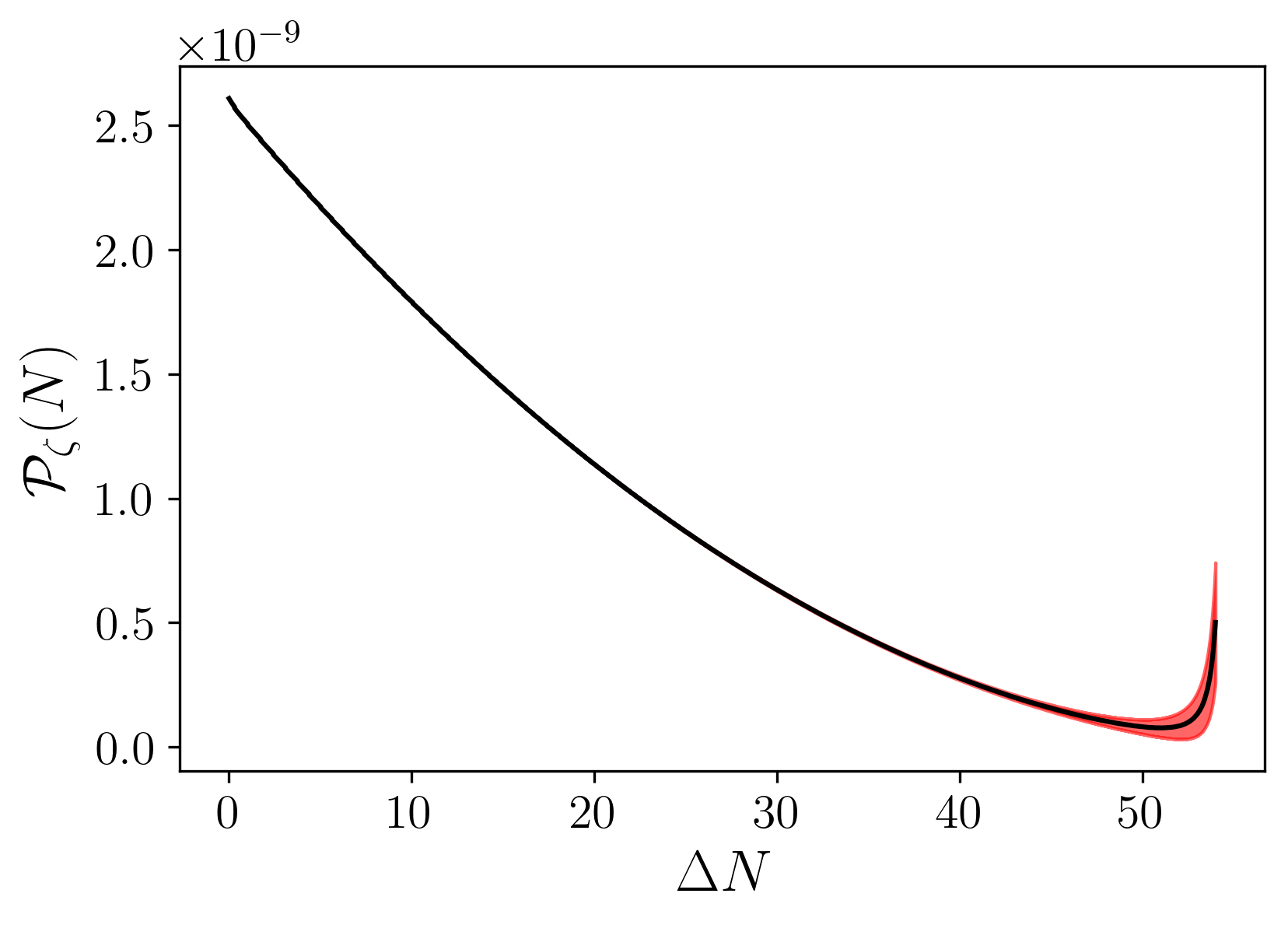}
\includegraphics[scale=0.8]{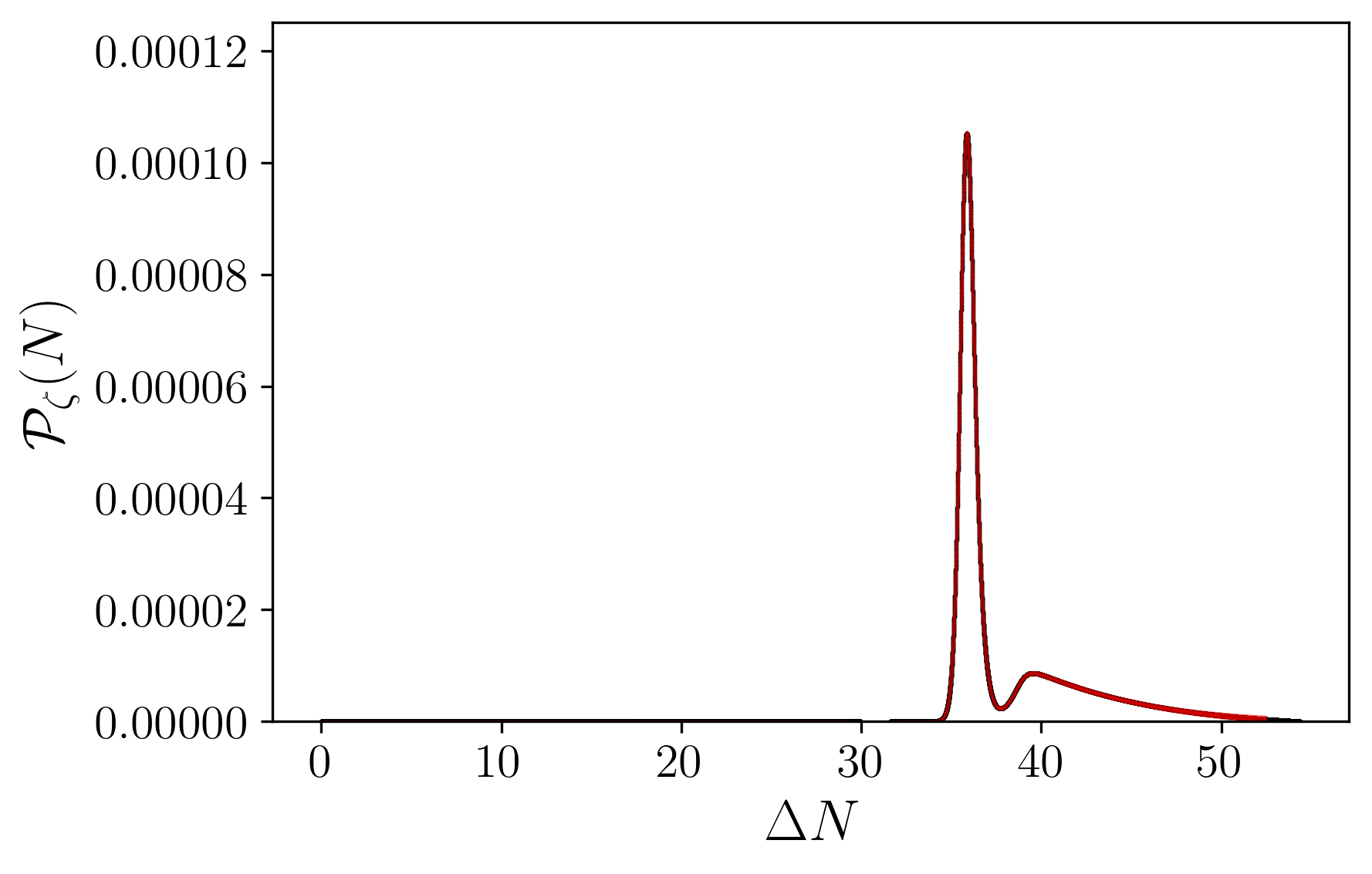}
\caption{Power spectrum of curvature perturbations for the chaotic potential (top panel) and deformed Starobinsky potential (bottom panel) in solid black with the red, shaded region showing the error bars obtained using jackknife sampling over $10^{7}$ realizations for $\sigma=0.01$.}
\label{fig:jackknife1}
\end{figure}

We obtain the error bars associated with the $\mathcal{P}_{\zeta}$ for both potentials after carrying out the jackknife resampling over $10^{6}$ and $10^{7}$ realizations of the stochastic process. In the top panel of Fig. \ref{fig:jackknife1} we plot the result for the chaotic potential. As expected, the error bars become more pronounced near the end of inflation because the fluctuations $\langle \delta\phi_{\text{st}}^{2} \rangle$ monotonically increase with $N$ (see the top subplot in Fig. \ref{fig:Adel}). The error bars on the deformed Starobinsky potential are more interesting. Comparisons can be made with Figs. \ref{fig:deformed_power} and \ref{fig:AdeltaPhi2_Deformed_Starobinsky}.

\section{Conclusions}

In this paper we numerically modeled stochastic inflationary dynamics under the influence of Gaussian white noise without any slow-roll simplifications. We studied two potentials: the quadratic potential and the deformed Starobinsky potential. The latter has an inflection point and the inflationary dynamics around such an inflection point is of interest to PBH formation. We were able to show that in the case of the deformed Starobinsky potential there is an amplification of the curvature power spectrum $\mathcal{P}_{\zeta}$ for modes that cross the horizon near the plateau region due to the interplay between the $\Xi_{\phi\phi}$ and $\Xi_{\pi\pi}$ noise terms. However, the amount of amplification was not as significant as was reported in \cite{Ezquiaga:2018gbw}. This has implications for inflationary model building aimed at PBH formation as discussed in Sec. \ref{sec:deformed_Starobinsky}. We have presented a detailed account of how to compute them by numerically solving the Mukhanov-Sasaki equations for $k_{\sigma}$ modes, where $\sigma$ is the coarse-graining scale, and incorporating them into the SDEs describing the evolution of the coarse-grained inflaton field and its conjugate momentum field. We conclude that the exact form of the stochastic noise terms have implications for the stochastic dynamics and subsequent computation of the curvature power spectrum as discussed ins Sec. \ref{sec:power}. The fact is that a slow-roll approximation of $\Xi_{\phi\phi}\simeq H^{2}/4\pi^2$ does not match numerical results even for completely slow-roll inflation models like $\phi^{2}$.\footnote{For example, at $\Delta N=0$, the exact numerical $\phi-\phi$ noise amplitude is around $4.5\times 10^{-11}$ whereas the analytical approximation is more closer to $4\times 10^{-11}$.} As a consequence, the SDEs are sourced by slightly larger random noise at each time step which is reflected in the final calculation of $\mathcal{P}_{\zeta}$ as a discrepancy in the CMB normalization of the curvature power spectrum.\\

Regardless of this, the numerical simulations did reveal the fact that the presence of a USR phase in the inflationary trajectory has important consequences with regards to amplifications to $\mathcal{P}_{\zeta}$. It is very well known by now that USR models require careful fine-tuning of the parameters (of which there are usually many) to not only produce the desired level of enhancement in the power spectrum, but also meet other criteria e.g., whether the amplification occurs for modes $k$ such that they collapse to form cosmologically significant PBHs. The stochastic calculations should potentially help in alleviating some of the difficulties associated with this extreme fine-tuning issue plaguing USR inflation models. However, for the deformed Starobinsky potential, we have identified a region at $\Delta N\sim 30$ where $\mathcal{P}_{\zeta}$ becomes negative due to $\langle \delta\phi_{\text{st}}^{2} \rangle$ being negatively sloped for a few $e$-folds around the region. This would coincide with the Mukhanov-Sasaki power spectrum if this decrease in $\langle \delta\phi_{\text{st}}^{2} \rangle$ was much more abrupt and confined to a very small band of $\Delta N$. This would have then corresponded to the downward facing cusp in $\mathcal{P}_{\zeta}$ between the slow-roll and USR periods which is a feature that is displayed by most, if not all, USR potentials.\\
\indent As a final comment, we also note that there were small changes in the height of the peak for the deformed Starobinsky potential for different values of the coarse-graining scale. For three separate values $\sigma=(5,7.5,10)\times 10^{-3}$, although the shape of $\mathcal{P}_{\zeta}$ remains similar, there are small changes in $\mathcal{P}_{\zeta}^{\text{max}}$. Of course the effects of smaller $\sigma$ can be explored. That would, however, require a larger number $e$-folds of background evolution beyond $N_{\star}$ such that $k_{\sigma}$ can be computed. Moreover, it would be of interest to study the effects of a Gaussian window function being used to define $\delta\hat{\phi}_{\bm{k}}$. This gives rise to colored noise where $\langle \xi_{\phi}(N)\xi_{\phi}(N') \rangle\sim\sech^{2}(N-N')$, which is left to be explored in a future project.

\section{Acknowledgements}
The authors thank Jose Mar\'{i}a Ezquiaga and Joseph Kapusta for online discussions and comments related to the project. Aritra De is supported by the U.S. DOE Grant No. DE-FG02- 87ER40328.

\section{Code availability}
The MATLAB codes developed for this project can be obtained from the authors upon request.

\appendix
\section{Derivation of Mukhanov-Sasaki equation}\label{sec:appendix_A}
Here we derive the Mukhanov-Sasaki equation that governs the evolution of the inflaton fluctuations. In a perturbed universe, along with the inflaton fluctuations, one must also deal with metric perturbations. Hence, to study how $\delta \phi$ evolves, we perturb the inflaton Klein-Gordon equation \cite{Mukhanov:2005sc,Riotto:2002yw}

\begin{equation}\label{eq:inflaton_KG}
\frac{1}{\sqrt{-g}}\partial_{\mu}\left( \sqrt{-g}g^{\mu\nu}\partial_{\nu}\phi \right)=\frac{\partial V}{\partial\phi}
\end{equation}
The metric tensor can be divided up into a background piece and perturbations using the Scalar-Vector-Tensor (SVT) decomposition. Generally, the line element takes the following form
\begin{equation}
ds^{2}=a(\eta)^{2}\left[ -(1+2\Phi)d\eta^{2} -2B_{i}dx^{i}d\eta +\left( (1-2\Psi)\delta_{ij}-2E_{ij} \right) \right]
\end{equation}
where $\Phi$ and $\Psi$ are scalar perturbations and $B_{i}$ and $E_{ij}$ are vector and tensor perturbations respectively. $E_{ij}$ also has the properties of being symmetric and traceless. Up to leading order in perturbations, the metric determinant can be expressed as
\begin{align*}
g&\simeq -(1+2\Phi)\text{det}\left[ a^{2}(1-2\Psi)+2a^{2}E_{ij} \right]\\
&= -a^{8}(1+2\Phi)(1-2\Psi +2E_{11})(1-2\Psi +2E_{22})(1-2\Psi +2E_{33})\\
&= -a^{8}(1+2\Phi-6\Psi)\numberthis
\end{align*}
The terms in $E_{ij}$ vanish due to the traceless property of the tensor. Hence
\begin{equation}
\sqrt{-g}\simeq a^{4}(1+\Phi - 3\Psi)
\end{equation}
This is then substituted into the inflaton Klein-Gordon equation Eq. \eqref{eq:inflaton_KG} to obtain
\begin{equation}
\phi''+2\mathcal{H}\phi'-\nabla^{2}\phi-(\Phi'+3\Psi'-\partial_{i}B_{i})\phi'=-a^{2}(1+2\Phi)\frac{\partial V}{\partial\phi}
\end{equation}
The primed quantities refer to derivatives with respect to the conformal time and $\mathcal{H}=a'/a$. Now, we further split the inflaton into the background and fluctuation parts $\phi(\eta,\bm{x})=\bar{\phi}(\eta)+\delta\phi(\eta,\bm{x})$. Once this is plugged into Eq. \eqref{eq:inflaton_KG}, there will be a homogeneous part $\bar{\phi}''+2\mathcal{H}\bar{\phi}'+\partial_{\phi}V$ that satisfies the background evolution and can be set to zero. The remaining terms describe the evolution of the inflaton fluctuations in a setting where the metric perturbations cannot be ignored.
\begin{equation}\label{eq:MS1}
\delta\phi''+2\mathcal{H}\delta\phi'-\nabla^{2}\delta\phi+a^{2}\partial_{\phi\phi}V\delta\phi=-2a^{2}\partial_{\phi}V\Phi+(\Phi'+3\Psi'+\nabla^{2}B)\bar{\phi}
\end{equation}

To proceed forward, one strategy that is adopted is to eliminate the metric perturbation variables from Eq. \eqref{eq:MS1} using the perturbed Einstein equations. From the perturbed Einstein equations, we get 
\begin{align*}
\tilde{\Psi}'+\mathcal{H}\Phi&=4\pi G \bar{\phi}'\delta\phi\\
\tilde{\Psi}'+\mathcal{H}\Phi +(-\mathcal{H}'+\mathcal{H}^{2})B&=4\pi G \left( \bar{\phi}'\delta\phi +\bar{\phi}'^{2}B \right)\\
(2\mathcal{H}'+\mathcal{H}^{2})\Phi + \mathcal{H}\Phi'+\tilde{\Psi}''+2\mathcal{H}\tilde{\Psi}'&=4\pi G\left[ (\bar{\phi}'\delta\phi'-\bar{\phi}'\Phi)-a^{2}\partial_{\phi}V\delta\phi \right] \numberthis
\end{align*}
where $\tilde{\Psi}=\Psi+\frac{1}{3}\nabla^{2}E$ is a new curvature variable. It is convenient to work in the spatially flat gauge where $\tilde{\Psi}_{Q}=0$ and $\Psi_{Q}=-\frac{1}{3}\nabla^{2}E_{Q}$. We introduce here the Mukhanov-Sasaki variable 
\begin{equation}
Q=\delta\phi +\frac{\bar{\phi}}{\mathcal{H}}\tilde{\Psi}
\end{equation}
Using the definition of $Q$, Eq. \eqref{eq:MS1} can be transformed into
\begin{equation}
Q''+2\mathcal{H}Q'-\nabla^{2}Q +a^{2}\partial_{\phi\phi}V Q=-2a^{2}\partial_{\phi}V\Phi_{Q}+\bar{\phi}'\Phi_{Q}'-\bar{\phi}'\nabla^{2}(E_{Q}'-B_{Q})
\end{equation}
Using one of the Einstein equations, we obtain $\Phi_{Q}=4\pi G\mathcal{H}^{-1}\bar{\phi}Q$. In order to eliminate the $\nabla^{2}(E_{Q}'-B_{Q})$ term, we recall that the gauge invariant Bardeen potentials\footnote{The Bardeen potentials $\Psi_{\text{GI}}$ and $\Phi_{\text{GI}}$ are the same when there are no anisotropic stresses present. This happens when $\delta T^{i}_{\;\;j}=0$ for $i\neq j$.} in the spatially flat gauge take the form

\begin{equation}
\Psi_{\text{GI}}=\Phi_{\text{GI}}=\mathcal{H}(E_{Q}'-B_{Q})
\end{equation}

The $\Phi_{\text{GI}}$ shows up in Eq. \eqref{eq:MS1} in the form of a Poisson equation $\nabla^{2}\Phi_{\text{GI}}=4\pi G a^{2}\delta\rho_{\text{GI}}$, where $\delta\rho_{\text{GI}}$ is a gauge invariant description of the perturbed energy density. One can show that the evolution of $Q$ can then be expressed in the following form

\begin{equation}\label{eq:MS2}
Q''+2\mathcal{H}Q'-\nabla^{2}Q +\left[a^{2}\partial_{\phi\phi}V -\frac{8\pi G}{a^{2}}\left( \frac{a^{2}}{\mathcal{H}}\bar{\phi}^{2} \right)'\right]Q=0
\end{equation}
The second term in the parenthesis is a consequence of the fact that metric perturbations are coupled to the inflaton fluctuations in a nontrivial manner through the Klein-Gordon equation. However, in slow-roll inflation, its effect is negligible. Equation \eqref{eq:MS2} can be expressed in $e$-fold time and Fourier decomposed to yield the expression in Eq. \eqref{eq:mukhanov_sasaki}.

\section{Evolution of coarse-grained fields}\label{sec:appendix_B}
Here we derive Eq. \eqref{eq:coarse_grain_evol} using the Hamiltonian framework. The equations will first be derived in conformal time and then converted to $e$-fold time. For simplicity, let us ignore perturbations of the spatial part of the metric, i.e. $g_{ij}=a(\eta)^{2}\delta_{ij}dx^{i}dx^{j}$. Considering the case when the inflaton is tightly coupled to metric perturbations, we get the following equations \cite{Grain:2017dqa}

\begin{align*}\label{eq:inflaton_with_metric}
\phi'&=\frac{1+\Phi}{a^{2}}\pi_{\phi}+\partial^{i}B\partial_{i}\phi\\
\pi_{\phi}'&=-a^{4}(1+\Phi)\partial_{\phi\phi}V +a^{2}\left[ \nabla^{2}\phi+\delta^{ij}\partial_{i}(\Phi\partial_{j}\phi) \right]+\partial_{i}\left[ (\partial^{i}B)\pi_{\phi} \right] \numberthis
\end{align*}
Now, the inflaton and its conjugate field is decomposed as follows $\phi=\bar{\phi}+\delta\hat{\phi}$ and $\pi_{\phi}=\bar{\pi}_{\phi}+\delta\hat{\pi}$, where
\begin{align*}
\delta\hat{\phi}(\eta,\bm{x})&=\int\frac{d^{3}k}{(2\pi)^{3/2}}W\left( \frac{k}{k_{\sigma}} \right)e^{-i\bm{k}\cdot\bm{x}}\hat{a}_{\bm{k}}\delta\phi_{\bm{k}}+\text{h.c.}\\
\delta\hat{\pi}(\eta,\bm{x})&=\int\frac{d^{3}k}{(2\pi)^{3/2}}W\left( \frac{k}{k_{\sigma}} \right)e^{-i\bm{k}\cdot\bm{x}}\hat{a}_{\bm{k}}\delta\pi_{\bm{k}}+\text{h.c.} \numberthis
\end{align*}
The field decompositions are then substituted into Eq. \eqref{eq:inflaton_with_metric} and linearized. Here spatial gradients of the fields $\bar{\phi}$ and $\bar{\pi}_{\phi}$ have been ignored. Also, we work in linear order in perturbations. After linearization
\begin{align*}
\bar{\phi}'&=\frac{1+\Phi}{a^{2}}\bar{\pi}_{\phi}+\frac{1}{a^{2}}\delta\hat{\pi}-\delta\hat{\phi}' \\
\bar{\pi}_{\phi}'&=-a^{4}(1+\Phi)\partial_{\phi}V-a^{4}\partial_{\phi\phi}V\delta\hat{\phi}-\delta\hat{\pi}'+a^{2}\nabla^{2}\delta\hat{\phi}+(\partial_{i}\partial^{i}B)\bar{\pi}_{\phi} \numberthis
\end{align*}

Now, the Fourier modes of the quantum fluctuations of $\delta\bar{\phi}$ and $\delta\bar{\pi}$ satisfy the following coupled differential equations\footnote{The metric perturbation terms $\Phi_{k}$ and $B_{k}$ can generally be ignored in cases where inflation is completely slow-roll. Then, one ends up with the following mode equation $$ \delta\phi_{k}''+2\mathcal{H}\delta\phi_{k}'+k^{2}\delta\phi_{k}+a^{2}\partial_{\phi\phi}V\delta\phi_{k}=0 $$}

\begin{align*}
\delta\phi_{k}'&=\frac{1}{a^{2}}\delta\pi_{k}+\frac{1}{a^{2}}\Phi_{k}\bar{\pi}_{\phi} \\
\delta\pi_{k}'&=-a^{4}\partial_{\phi\phi}V\delta\phi_{k}-a^{2}k^{2}\delta\phi_{k}-a^{4}\Phi_{k}\partial_{\phi}V-k^{2}B_{k}\bar{\pi}_{\phi} \numberthis
\end{align*}
and it can be verified that these coupled differential equations are in fact the Mukhanov-Sasaki equation in Eq. \eqref{eq:MS2}. It is very easy to verify that the coarse-grained fields evolve under the following differential equations

\begin{align*}
\bar{\phi}'&=\frac{\bar{\pi}_{\phi}}{a^{2}}+\xi_{\phi}\\
\bar{\pi}_{\phi}'&=-a^{4}\partial_{\phi}V+\xi_{\pi}\numberthis
\end{align*}
where $\xi_{\phi}$ and $\xi_{\pi}$ are the noise terms associated with the inflaton field and its conjugate defined as
\begin{align*}
\xi_{\phi}&=-\int\frac{d^{3}k}{(2\pi)^{3/2}}W'\left( \frac{k}{k_{\sigma}} \right)e^{-i\bm{k}\cdot\bm{x}}\hat{a}_{k}\delta\phi_{k}+\text{h.c.}\\
\xi_{\pi}&=-\int\frac{d^{3}k}{(2\pi)^{3/2}}W'\left( \frac{k}{k_{\sigma}} \right)e^{-i\bm{k}\cdot\bm{x}}\hat{a}_{k}\delta\pi_{k}+\text{h.c.}\numberthis
\end{align*}
It is straightforward to transform these equations from conformal time to $e$-fold time as is done in the main body of the paper. The correlations of the noise terms can then be calculated by considering the time-ordered vacuum expectation values of the $\xi$ terms, which are technically still quantum operators. However, in the superhorizon, large squeezing limit, the following correspondance can be established
\begin{equation}
\langle \xi_{i}(\eta,\bm{x})\xi_{j}(\eta,\bm{x}) \rangle=\bra{0}T\left[ \hat{\xi}_{i}(\eta,\bm{x})\hat{\xi}_{j}(\eta,\bm{x}) \right]\ket{0}
\end{equation}
where the subscripts $i,j$ refer to $\phi$ and $\pi_{\phi}$ respectively and $T$ stands for time-ordering. A detailed derivation of this can be found in \cite{Vennin:2014lfa}.

\section{Correlation functions}\label{sec:correlation_FP}

Here we derive the expression of the power spectrum given by Eq. \eqref{eq:P_zeta_approx} starting from the Fokker-Planck equation. Labelling the coarse-grained inflaton and conjugate momentum fields with the vector $\bm{\Phi}=(\bar{\phi},\bar{\pi}_{\phi})$, the Fokker-Planck equation can be written as
\begin{equation}
\frac{\partial}{\partial N}P(N;\bm{\Phi})=-\frac{\partial}{\partial\bm{\Phi}_{A}}(D_{A}P(N;\bm{\Phi}))+\frac{1}{2}\Xi_{AB}\frac{\partial^{2}}{\partial\bm{\Phi}_{A}\partial\bm{\Phi}_{B}}P(N;\bm{\Phi})
\end{equation}
where $D_{\phi}$ and $D_{\pi}$ are the drift components given by
\begin{align}
D_{\phi}&=\bar{\pi}_{\phi} \nonumber\\
D_{\pi}&=-(3-\epsilon_{1})\left( \bar{\pi}_{\phi}+\frac{\partial_{\phi}V}{V} \right) 
\end{align}
Defining the stochastic fluctuations as $\delta\phi=\bar{\phi}-\phi_{c}$ and $\delta\pi=\bar{\pi}_{\phi}-\pi_{c}$, the PDF can be used to describe the statistical moments of the fluctuations.
\begin{equation}
 \langle \delta\phi^{n}\delta\pi^{m} \rangle=\int d\bar{\pi}_{\phi}\int d\bar{\phi}\;(\bar{\phi}-\phi_{c})^{n}(\bar{\pi}_{\phi}-\pi_{c})^{m}P(N;\bar{\phi},\bar{\pi}_{\phi}) 
 \end{equation}
Now we compute the derivative of $\langle \delta\phi^{n}\delta\pi^{m} \rangle$ with respect to $N$. Taking a derivative with respect to $e$-folds, we get
\begin{align}
\frac{d}{dN}\langle \delta\phi^{n}\delta\pi^{m} \rangle&=\int d\bar{\pi}_{\phi}\int d\phi\bigg[ n\delta\phi^{n-1}\left( \frac{d\bar{\phi}}{dN}-\frac{d\phi_{c}}{dN} \right)\delta\pi^{m}P \nonumber\\
\qquad & \quad\quad\quad+ m\delta\phi^{n}\delta\pi^{m-1}\left( \frac{d\bar{\pi}_{\phi}}{dN}-\frac{d\pi_{\phi,c}}{dN} \right)P + \delta\phi^{n}\delta\pi^{m}\frac{\partial P}{\partial N}\bigg] \nonumber\\
&=\int d\bar{\pi}_{\phi}\int d\bar{\phi}\bigg[n\delta\phi^{n-1}\delta\pi^{m}P(D_{\phi}-D_{\phi}^{c})+m\delta\phi^{n}\delta\pi^{m-1}P(D_{\pi}-D_{\pi}^{c})\bigg]
\qquad & \quad\quad\quad\quad\quad\quad\quad\quad+ \delta\phi^{n}\delta\pi^{m}\frac{\partial P}{\partial N}\bigg] \nonumber\\
&=n(\langle \delta\phi^{n-1}\delta\pi^{m}D_{\phi} \rangle-\langle \delta\phi^{n-1}\delta\pi^{m}D_{\phi}^{c} \rangle)+m(\langle \delta\phi^{n}\delta\pi^{m-1}D_{\pi} \rangle-\langle \delta\phi^{n}\delta\pi^{m-1}D_{\pi}^{c} \rangle) \nonumber\\
\qquad &\quad\quad\quad\quad\quad\quad\quad\quad +\int d\bar{\pi}_{\phi}\int d\bar{\phi}\delta\phi^{n}\delta\pi^{m}\frac{\partial P}{\partial N} 
\end{align}
where $D_{\phi}^{c}$ and $D_{\pi}^{c}$ are the drift terms evaluated for the classical, non-stochastic fields. When the term containing the partial derivative of the PDF is expanded out using the Fokker-Planck equation, six terms will be produced when the indices $A$ and $B$ are summed over. To simplify this, we can focus on the following term in the expression.
\begin{equation}
\int d\bar{\pi}_{\phi}\int d\bar{\phi}\;\delta\phi^{n}\delta\phi^{m}\frac{\Xi_{\phi\phi}}{2}\frac{\partial^{2}P}{\partial\bar{\phi}^{2}} \nonumber
\end{equation}
The derivatives acting on $P$ can be removed by performing integration by parts two times at the expense of picking up two surface terms.
\begin{align}
\int d\bar{\pi}_{\phi}\;\delta\phi^{n}\delta\pi^{m}\frac{\Xi_{\phi\phi}}{2}\frac{\partial P}{\partial\bar{\phi}}\bigg\lvert_{-\infty}^{+\infty}-&n\int d\bar{\pi}_{\phi}\;\delta\phi^{n-1}\delta\pi^{m}\frac{\Xi_{\phi\phi}}{2}P\big\lvert_{-\infty}^{+\infty}
 &  +n(n-1)\int d\bar{\pi}_{\phi}\int d\bar{\phi}\;\delta\phi^{n-2}\delta\pi^{m}\frac{\Xi_{\phi\phi}}{2}P \nonumber
\end{align}
The limits $\pm\infty$ on the PDF and its derivative is arbitrary at this point. However, one can always consider a UV cutoff for the fields such that they cannot explore arbitrarily large field values by imposing some reflective boundary conditions. Using the conservation of probability, we can consider the PDF as decaying to zero at the end points of field space. Additionally, we can impose the condition that the probability flux vanishes at the endpoints. These conditions imply that
\begin{equation}
P(N;\bar{\phi})\big\lvert^{+\infty}_{-\infty}=\frac{\partial}{\partial \bar{\phi}}P(N;\bar{\phi})\bigg\lvert_{-\infty}^{+\infty}=0
\end{equation}
Imposing these conditions on the rest of the terms, the derivative of the moments simplifies to
\begin{align}\label{eq:correlation_derivative}
\frac{d}{dN}\langle \delta\phi^{n}\delta\pi^{m} \rangle&=n(\langle \delta\phi^{n-1}\delta\pi^{m}D_{\phi} \rangle-\langle \delta\phi^{n-1}\delta\pi^{m}D_{\phi}^{c} \rangle)+m(\langle \delta\phi^{n}\delta\pi^{m-1}D_{\pi} \rangle-\langle \delta\phi^{n}\delta\pi^{m-1}D_{\pi}^{c} \rangle) \nonumber \\
\qquad & +\frac{1}{2}n(n-1)\langle \delta\phi^{n-2}\delta\pi^{m} \rangle\Xi_{\phi\phi}+\frac{1}{2}m(m-1)\langle \delta\phi^{n}\delta\pi^{m-2} \rangle\Xi_{\phi\phi}\nonumber \\
\qquad & +\frac{1}{2}nm\langle \delta\phi^{n-1}\delta\pi^{m-1} \rangle(\Xi_{\phi\pi}+\Xi_{\pi\phi})
\end{align}

To extract some meaningful information, the drift terms need to be expanded. The drift terms are first expressed as
\begin{align}
D_{\phi}&=\delta\pi+\pi_{c}\nonumber \\
D_{\pi}&=-\left( 3-\frac{1}{2}(\delta\pi+\pi_{c})^{2} \right)\left[(\delta\pi+\pi_{c})-\frac{\partial_{\phi}V}{V}\bigg\lvert_{\phi_{c}+\delta\phi}\right]
\end{align}
Substituting these expressions into Eq. \eqref{eq:correlation_derivative}, the terms containing $D_{\phi}$, $D_{\pi}$ and their classical counterparts produce the following upon multiple cancellations
\begin{align}
&n(\langle \delta\phi^{n-1}\delta\pi^{m}D_{\phi} \rangle-\langle \delta\phi^{n-1}\delta\pi^{m}D_{\phi}^{c} \rangle)+m(\langle \delta\phi^{n}\delta\pi^{m-1}D_{\pi} \rangle-\langle \delta\phi^{n}\delta\pi^{m-1}D_{\pi}^{c} \rangle)\nonumber \\
&=n\langle \delta\phi^{n-1}\delta\pi^{m+1} \rangle-3m\langle \delta\phi^{n}\delta\pi^{m} \rangle+\frac{3}{2}m\langle \delta\phi^{n}\delta\pi^{m+1} \rangle\pi_{c}+\frac{3}{2}m\langle \delta\phi^{n}\delta\pi^{m} \rangle\pi^{2}_{c}\nonumber\\
&+\frac{1}{2}m\langle \delta\phi^{n}\delta\pi^{m+2} \rangle-\frac{m}{2}\bigg\langle \delta\phi^{n}\delta\pi^{m+1}\frac{\partial_{\phi}V}{V}\bigg\lvert_{\phi_{c}+\delta\phi} \bigg\rangle-m\bigg\langle \delta\phi^{n}\delta\pi^{m}\frac{\partial_{\phi}V}{V}\bigg\lvert_{\phi_{c}+\delta\phi} \bigg\rangle\pi_{c}
\end{align}
The terms with the partial derivative of the potential are evaluated at $\phi_{c}+\delta\phi$ and can be simplified by consider a Taylor expansion. If $f$ is the function of the fields, then
\begin{equation}
f(\phi_{c}+\delta\phi)=f(\phi_{c})+f^{(1)}\delta\phi+\frac{1}{2!}f^{(2)}\delta\phi^{2}+\cdot\cdot\cdot=f(\phi_{c})+\sum_{k=1}^{\infty}\frac{f^{(k)}}{k!}\delta\phi^{k}
\end{equation}
Now, since $f=\partial_{\phi}/V=(\ln V)_{,\phi}$, the relevant expansion would produce
\begin{equation}
\frac{\partial_{\phi}V}{V}\bigg\lvert_{\phi_{c}+\delta\phi}=\frac{\partial_{\phi}V}{V}\bigg\lvert_{\phi_{c}}+\sum_{k=1}^{\infty}\frac{(\ln V)^{(k+1)}}{k!}\delta\phi^{k}
\end{equation}
Then,
\begin{align}\label{eq:correlation_derivative}
\frac{d}{dN}\langle \delta\phi^{n}\delta^{m} \rangle&=n\langle \delta\phi^{n-1}\delta\pi^{m+1} \rangle-3m\langle \delta\phi^{n}\delta\pi^{m}\rangle\nonumber\\
&+\frac{m}{2}(\langle \delta\phi^{n}\delta\pi^{m+2}\rangle+3\langle \delta\phi^{n}\delta\pi^{m+1}\rangle\pi_{c}+3\langle \delta\phi^{n}\delta\pi^{m}  \rangle\pi_{c}^{2} )\nonumber\\
&+m(3-\epsilon_{1})\sum_{k=1}^{\infty}\frac{(\ln V)^{(k+1)}}{k!}\langle \delta\phi^{m+k}\delta\pi^{m-1} \rangle\nonumber \\
&-\frac{m}{2}\left( \langle \delta\phi^{n}\delta\pi^{m+1}\frac{\partial_{\phi}V}{V} \rangle+2\langle \delta\phi^{n}\delta\pi^{m}\frac{\partial_{\phi}V}{V} \rangle\pi_{c} \right)\nonumber\\
&+\frac{1}{2}n(n-1)\langle \delta\phi^{n-2}\delta\pi^{m} \rangle\Xi_{\phi\phi}+\frac{1}{2}m(m-1)\langle \delta\phi^{n}\delta\pi^{m-2} \rangle\Xi_{\pi\pi}\nonumber\\
&+\frac{1}{2}nm\langle \delta\phi^{n-1}\delta\pi^{m-1} \rangle(\Xi_{\phi\pi}+\Xi_{\pi\phi})
\end{align}
The computation of the derivative of the statistical moments therefore involves truncating the infinite sum to the required values of $n$ and $m$. A good way of doing this would be to truncate $\sum_{k}(\ln V)^{(k+1)}/k!$ at order $n+m$. For $n=2$ and $m=0$, we are able to reproduce Eq. \eqref{eq:P_zeta_approx}.

\bibliography{stochastic_inflation_manuscript}

\end{document}